\documentclass[fleqn,usenatbib]{mnras}

\usepackage{newtxtext,newtxmath}

\usepackage[T1]{fontenc}
\DeclareRobustCommand{\VAN}[3]{#2}
\let\VANthebibliography\thebibliography
\def\thebibliography{\DeclareRobustCommand{\VAN}[3]{##3}\VANthebibliography}

\usepackage{graphicx}	
\usepackage{amsmath}	

\title[A Be star is born]{Birth of a Be star: an APOGEE search for Be stars forming through binary mass transfer}

\author[El-Badry et al.]{Kareem El-Badry,$^{1,2,3,4}$\thanks{E-mail: kareem.el-badry@cfa.harvard.edu}
Charlie Conroy,$^{1}$
Eliot Quataert,$^{4,5}$
Hans-Walter Rix,$^{3}$ \newauthor
Jonathan Labadie-Bartz,$^{6}$
Tharindu Jayasinghe,$^{7,8}$ 
Todd Thompson,$^{7,8}$
Phillip Cargile,$^{1}$
Keivan G.\ Stassun,$^9$
\newauthor
Ilya Ilyin$^{10}$
\\
$^{1}$Center for Astrophysics $|$ Harvard \& Smithsonian, 60 Garden Street, Cambridge, MA 02138, USA\\
$^{2}$Harvard Society of Fellows, 78 Mount Auburn Street, Cambridge, MA 02138\\
$^{3}$Max-Planck Institute for Astronomy, K\"onigstuhl 17, D-69117 Heidelberg, Germany\\
$^{4}$Department of Astronomy, University of California, Berkeley, Berkeley, CA, 94720, USA\\
$^{5}$Department of Astrophysical Sciences, Princeton University, Princeton, NJ 08544, USA\\
$^{6}$Instituto de Astronomia, Geof\'{i}sica e Ci\^{e}ncias Atmosf\'{e}ricas, Universidade de S\~{a}o Paulo, Rua do Mat\~{a}o 1226, Cidade Universit\'{a}ria, 05508-900 S\~{a}o Paulo,  Brazil \\
$^{7}$Department of Astronomy, The Ohio State University, 140 West 18th Avenue, Columbus, OH 43210, USA\\
$^{8}$Center for Cosmology and Astroparticle Physics, The Ohio State University, 191 W. Woodruff Avenue, Columbus, OH 43210, USA \\
$^{9}$Department of Physics and Astronomy, Vanderbilt University, Nashville, TN 37235, USA \\
$^{10}$Leibniz Institute for Astrophysics Potsdam (AIP), An der Sternwarte 16, D-14482 Potsdam, Germany
}

\date{Accepted to MNRAS}

\pubyear{2022}

\begin{document}
\label{firstpage}
\pagerange{\pageref{firstpage}--\pageref{lastpage}}
\maketitle

\begin{abstract}
Motivated by recent suggestions that many Be stars form through binary mass transfer, we searched the APOGEE survey for Be stars with bloated, stripped companions. From a well-defined parent sample of 297 Be stars, we identified one mass-transfer binary, HD 15124. The object consists of a main-sequence Be star ($M_{\rm Be}=5.3\pm 0.6 \,M_{\odot}$) with a low-mass ($M_{\rm donor}=0.92\pm 0.22\,M_{\odot}$), subgiant companion on a 5.47-day orbit. The emission lines originate in an accretion disk caused by ongoing mass transfer, not from a decretion disk as in classical Be stars. Both stars have surface abundances bearing imprint of CNO processing in the donor's core: the surface helium fraction is $Y_{\rm He}\approx 0.6$, and the nitrogen-to-carbon ratio is 1000 times the solar value. The system's properties are well-matched by binary evolution models in which mass transfer begins while a $3-5\,M_{\odot}$ donor leaves the main sequence, with the originally less massive component becoming the Be star. These models predict that the system will soon become a detached Be + stripped star binary like HR 6819 and LB-1, with the stripped donor eventually contracting to become a core helium-burning sdO/B star. Discovery of one object in this short-lived ($\sim$1\,Myr) evolutionary phase implies the existence of many more that have already passed through it and are now Be + sdO/B binaries. We infer that $(10-60)\,\%$ of Be stars have stripped companions, most of which are $\sim 100\,\times$ fainter than the Be stars in the optical. Together with the dearth of main-sequence companions to Be stars and recent discovery of numerous Be + sdO/B binaries in the UV, our results imply that binarity plays an important role in the formation of Be stars.
\end{abstract}




\section{Introduction}

Binary mass transfer underlies many of the most important open questions in astrophysics. A large fraction of all stars -- particularly massive stars -- are in binaries that exchange mass at some point in their evolution \citep[e.g.][]{Sana2012, Kobulnicky2014, Moe2017}, and mass transfer often dramatically changes the lives and deaths of stars \citep[e.g.,][]{deMink2013}. Realistic modeling of interacting binaries is one of the major areas of research in modeling of unresolved stellar populations \citep[e.g.][]{Eldridge2017}. Mass transfer is challenging to study observationally because it often occurs on short timescales, with most observed binaries found before or after periods of mass transfer. Thus, although stellar population synthesis models that include binaries make significantly different predictions from single-star models for the ionizing flux, metal enrichment, and supernova rates of unresolved stellar populations \citep[e.g.][]{Stanway2016, Ma2016, Steidel2016}, the underlying binary evolution models have been subject to only limited tests with observations of individual binaries in the nearby Universe \citep[e.g.][]{Han2003, Sana2013, Sen2021}. 

A somewhat counterintuitive but common outcome of binary evolution with mass transfer is mass ratio inversion \citep[e.g.][]{Giuricin1983}, wherein the less-massive component of a binary is observed to be the more evolved star. Such ``Algol-type'' binaries form when the initially more massive component of a binary loses a large fraction of its mass to a companion \citep[e.g.][]{Paczynski1971}, generally through stable (i.e., long-lived and non-dynamical) mass transfer.
Explaining the existence of Algol-type binaries as a consequence of mass transfer was one of the great achievements of stellar evolution modeling in the last century \citep[see][for a review]{Pustylnik1998}. 

Although stably mass-transferring binaries have been  studied intensely for the better part of a century, several recent developments have renewed interest in their evolution. The first is the mounting evidence that spin-up through mass transfer is among the primary causes of the ``Be'' phenomenon; i.e., the prevalence of rapidly-rotating B stars with emission lines, which usually trace circumstellar disks. The idea that Be stars might be binary evolution products spun up by mass transfer, first popularized by \citet{Pols1991}, has recently been bolstered by the discovery in the UV of more than a dozen, compact core-helium burning stars (``sdO/B'' stars) as companions to Be stars \citep[e.g.][]{Wang2018, Wang2021}. Some of these companions have also been confirmed interferometrically \citep{Klement2021_interferometry}. Other indirect lines of evidence have also recently suggested that many Be stars have hard-to-detect companions stripped by binary interactions: the frequency of {\it main-sequence} companions to Be stars is unusually low \citep{Bodensteiner2020b}, and the SEDs of many Be star disks hint at tidal truncation by a companion \citep{Klement2019}. Several lines of evidence also  suggest that sdO/B stars are essentially always formed via binary interactions \citep[e.g.][]{Han2003, Pelisoli2020, Geier2022}, so studying the formation channel of Be stars can also shed light onto the formation of sdO/B stars.

Second, several unusual binaries have recently been identified that most likely contain a Be star and a warm ($T_{\rm eff}\approx 15,000\,\rm K$), bloated ($R\approx 5\,R_{\odot}$) companion \citep[][]{Liu2019, Rivinius2020, Saracino2021}. All of these systems were initially interpreted as containing a main-sequence B star orbiting a black hole, but subsequent studies \citep[e.g.][]{Irrgang2020, Shenar2020, Bodensteiner2020, El-Badry2021, El-Badry_burdge2021} have shown that the B stars are not on the main sequence, but appear to be undermassive ($\approx 0.5-1.5\,M_{\odot}$) stripped products of binary evolution that are currently contracting to become core helium burning sdO/B stars. Because the current evolutionary state of these objects is short-lived, their discovery implies that there may be many other systems with similar evolutionary histories yet to be discovered. These  Be + sdO/B binaries would be long-lived but challenging to identify: the Be star usually contributes the vast majority of the total light in the optical, and the predicted few-$\rm km\,s^{-1}$ radial velocity (RV) shifts of the Be star  are only marginally detectable due to Be stars' large rotation velocities and disk-driven spectral variability.

The chances of detecting compact Be + sdO/B systems are higher in the UV, where  the hot sdO/B star contributes a larger fraction of the total light. Even so, only the tail of the population with the hottest sdO/B stars are unambiguously detectable \citep[e.g.][]{Schootemeijer2018}. Compared to typical sdO/B stars, a majority of the sdO/B companions to Be stars discovered in the UV \citep[e.g.][]{Gies_1998, Koubsky_2012, Peters_2013, Wang2018, Wang2021} are unusually hot and luminous. This is at least partly a selection effect, since companions are more likely to be detectable when they contribute a larger fraction of the total light.  

While sdO/B companions are most easily detected in the UV, the optical and IR wavelengths are optimal for characterizing earlier evolutionary stages of systems that will later become Be + sdO binaries. For instance, shortly after leaving the main sequence, the initially more massive star can be relatively luminous and rich in spectral features in the optical and/or IR regions, and all but invisible in the UV. Surveys in different wavelength regimes are thus necessary to build a more complete picture of the evolutionary paths that can lead to rapidly rotating Be stars and their stripped companions. 

In this paper, we further test the hypothesis that binary mass transfer is an important formation channel for Be stars by conducting a systematic search for Be stars in binaries being spun up by mass transfer. This allows us to then obtain a more robust estimate of the fraction of Be stars that have gone through a mass-transfer phase than could be obtained from HR 6819 and LB-1, as both systems were identified serendipitously from ill-defined effective parent samples. The search yields one binary, HD 15124, on which most of the paper is focused.

The remainder of this paper is organized as follows. Section~\ref{sec:search} describes our search for Be + stripped star binaries in APOGEE. Section~\ref{sec:data} presents data on HD 15124, including the APOGEE spectra, follow-up optical spectra, light curves, radial velocities, and the broadband SED. Section~\ref{sec:stellar_params} summarizes our constraints on the physical parameters of the binary. In Section~\ref{sec:mesa}, we construct binary evolution models to investigate the system's formation history and predicted future evolution. We discuss the implications of our results for the broader Be star population in Section~\ref{sec:discussion}. The paper's main results are summarized in Section~\ref{sec:conclusions}, and the Appendices provide more information on several aspects of our modeling.

\section{APOGEE search for Be stars with narrow-lined companions}
\label{sec:search}

\begin{figure*}
    \centering
    \includegraphics[width=\textwidth]{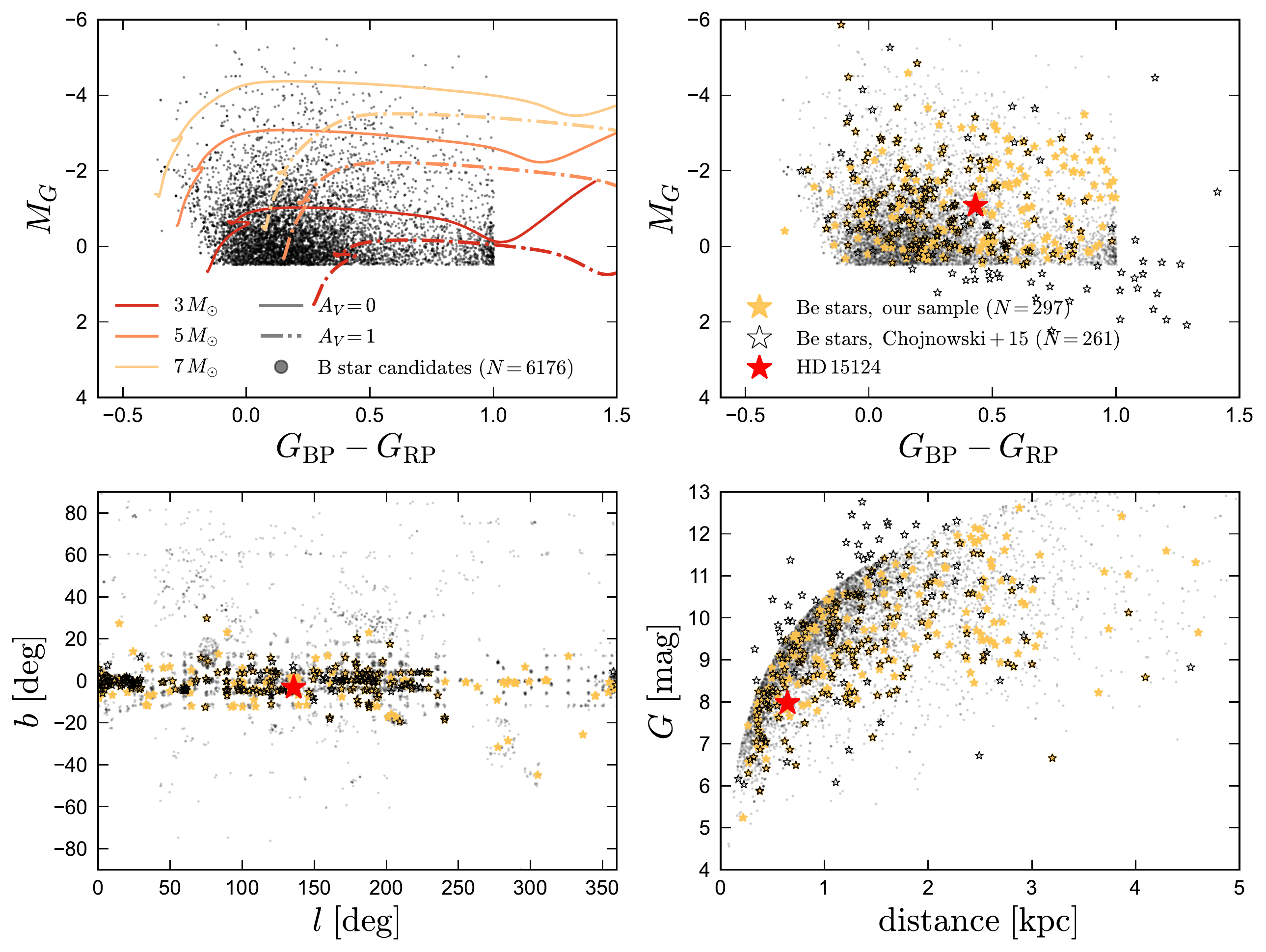}
    \caption{Sample of candidate Be stars in context. Upper left: Color-magnitude diagram. Gray points show APOGEE DR16 sources with $G_{\rm BP}-G_{\rm RP} <1$ and $M_{G} < 0.5$; i.e., hot and luminous stars. Colored lines show MIST evolutionary tracks without extinction (solid) and with $A_V=1$ (dot-dashed). Upper right: We inspected the APOGEE spectra of the 6176 sources and identified 297 Be stars with significant emission in the Brackett series (gold stars). Open stars show another sample of Be stars identified from APOGEE spectra by \citet{Chojnowski2015}, a majority of which overlap with our sample and a few of which fall outside our CMD selection region. 
    Red star shows HD 15124. Lower left: sky distribution. Most Be stars, including HD 15124, are in the Galactic plane. Lower right: apparent magnitude and distance. HD 15124 is in the nearest and brightest quartile of the Be star sample. }
    \label{fig:cmd_selection}
\end{figure*}

We began by searching DR16 of the APOGEE survey \citep{Majewski2017, Jonsson2020} for Be stars with narrow-lined companions. Although the primary goals of the APOGEE survey are centered on cool giants, it also observed several thousand OB stars all across the sky, mainly for use as telluric standards.

\subsection{Identifying B and Be stars}
 \label{sec:Be_search}
As a first step, we selected hot and luminous stars observed by APOGEE in the {\it Gaia} color-magnitude diagram (CMD). Of the 437,445 stars included in DR16, 432,541 had a corresponding  source (as identified in the \texttt{allStar} table) in {\it Gaia} DR2. 339,374 of these had at least a marginally significant parallax (\texttt{parallax\_over\_error} > 3), and 338,733 also had a reported \texttt{bp\_rp} color. From this sample, we selected candidate OB stars as those satisfying $M_G < 0.5$, where $M_G = G + 5\log \left (\varpi/100\right)$, and \texttt{bp\_rp < 1} (gray points in Figure~\ref{fig:cmd_selection}). This left us with 6,176 hot and luminous star candidates with APOGEE spectra. 5,295 of these were observed at APO, 1,018 were observed at LCO, and 359 were observed at both sites.

Colored lines in Figure~\ref{fig:cmd_selection} show synthetic photometry for single-star MIST evolutionary models \citep{Choi2016} with $A_V=0$ (solid) and $A_V=1$ (dot-dashed). These show that the sample is expected to contain sources with masses $M\gtrsim 3\,M_{\odot}$ and moderate extinction.  For example, a typical $5\,M_{\odot}$ main-sequence star with $T_{\rm eff}= 16\,000$\,K and $\log g = 4$ would have $M_{G}\approx -1$ and $G_{\rm BP}-G_{\rm RP}\approx -0.2$ with no extinction, and would thus fall within our selection region as long as $A_V\lesssim 1.75$. This CMD-based selection is crude, and more careful selections exist in the literature \citep[e.g.][]{Zari2021}. Our goal was to cast a wide net and include a majority of hot and luminous stars in our sample, removing cooler stars later. 

We retrieved the combined spectra (\texttt{apStar}/\texttt{asStar} files) of all 6,176  hot and luminous star candidates and inspected them individually to identify Be stars. We focused on the Brackett 11$\to$4 line at 16,802\,\AA, which is usually the strongest emission line in Be stars within the APOGEE wavelength coverage \citep{Chojnowski2015}. This search yielded 297 Be stars with unambiguous emission in the Brackett lines; these are shown in Figure~\ref{fig:cmd_selection} with gold stars. For comparison, we also show the sample of 261 Be star candidates identified from APOGEE DR12  by \citet{Chojnowski2015} as open star symbols. Examples of several spectra are shown in Figure~\ref{fig:example_spectra}. 
 
146 of the Be stars in our sample are also in the \citet{Chojnowski2015} sample. A majority of our new identifications are stars observed after DR12. Of the 115 Be stars identified by \citet{Chojnowski2015} that are {\it not} included in our sample, 18 are not included in the DR16 \texttt{allStar} table, and 16 did not pass our {\it Gaia} \texttt{parallax\_over\_error} cut. 36 passed other cuts, but fell outside our CMD selection (left panel of Figure~\ref{fig:cmd_selection}). Finally, 45 fell within our selection region in the CMD, but were not identified as Be stars in our visual inspection. We re-inspected the spectra of these 45 stars and found that most of them show weak or ambiguous emission (e.g., bottom panel of Figure~\ref{fig:example_spectra}) and therefore were not flagged in our search.

As is evident from the upper right panel of Figure~\ref{fig:cmd_selection}, 14\% of the Be stars from the \citet{Chojnowski2015} sample fall outside our CMD selection region. Most of these sources have high extinctions. This suggest that we could have obtained a 10-20\% larger sample of Be stars by expanding our CMD selection region. We opted for a more restrictive selection because a larger CMD search region would lead to more contamination from giants and lower-mass main sequence stars, significantly increasing the number of spectra requiring visual inspection.

In summary, the parent sample within which we search for bloated stripped companions contains 297 Be stars. The sample is bright (median apparent magnitude $G=9.4$) and relatively nearby (median distance 1.3 kpc). Most stars have precise parallaxes (median \texttt{parallax\_over\_error} = 33 with {\it Gaia} eDR3 data). After correcting for extinction and reddening using the 3D dust map from \citet{Green2019}, the median color and absolute magnitude of the sample is $\left({\rm bp\_rp}\right)_{0}=-0.07$ and $M_{G,0}=-1.6$. This corresponds to spectral type B3; i.e., $\approx 5\,M_{\odot}$ for stars near the main sequence. This is a slightly lower mass than the spectral type where the frequency of Be stars peaks (B1.5-B2; e.g. \citealt{Kogure1982, Rivinius2013}), reflecting the fact that lower-mass stars are longer-lived and favored by the IMF. 

\subsection{Search for Be stars with narrow lines and/or RV variability}
\label{sec:visit-inspect}
For each of the 297 Be stars, we retrieved and inspected the individual-visit spectra (\texttt{apVisit/asVisit} files) for evidence of binarity. The median number of visits per target in DR16 is 5. All but a few targets have at least 3 visits, and about a third have more than 7. The median time baseline spanned by the observations for a single target is 295 days, and the median single-visit SNR is 208 per pixel.

The primary features we searched for were: 
\begin{enumerate}
    \item {\it Epoch-to-epoch RV variability}: For low-mass companions with periods $\lesssim 1$ year, as expected for binaries with ongoing or recently terminated mass transfer, we expect the companions to have typical observable epoch-to-epoch RV shifts of $\gtrsim 10\,\rm km\,s^{-1}$. The RV shifts of the Be star, on the other hand, are generally not expected to be detectable, due to the Be stars' large masses and rapid rotation rates.
    We found that the RVs reported in the \texttt{allVisit} table are often unreliable for Be stars, so we did not use these in our analysis. Instead, we searched for objects in which there appeared to be a genuine Doppler shift of photospheric absorption lines after barycentric correction.

    \item  {\it Narrow absorption lines}: Be stars are expected to be rapidly rotating, and thus, to have rotationally smeared-out absorption lines except when viewed nearly pole-on. In contrast, the bloated, stripped companions we seek to identify are expected to have been tidally synchronized by mass transfer, and thus to have relatively low projected rotation velocities. We thus regarded the presence of narrow absorption lines as a promising indicator of a binary companion.

    \item {\it Line profile variability}: We also searched for epoch-to-epoch changes in emission line profile shapes, which could also be related to binarity. Such changes are, however, ubiquitous in the emission lines of Be stars \citep[e.g.][]{Slettebak1979, Dachs1981, Okazaki1991} and are thought to be due primarily to changes in the structure of the circumstellar disk, often unrelated to binarity. We therefore focused primarily on absorption lines, and did not consider emission line profile changes alone to be strong evidence for binarity.
    
\end{enumerate}

\subsubsection{Search sensitivity}
The initial motivation for our search was to identify relatively hot, bloated companions similar to those discovered in LB-1 and HR 6819. The stripped stars in these objects have already terminated mass transfer and begun contraction toward the extreme horizontal branch, with effective temperatures of 13,000 and 16,000 K.  

Unfortunately, stars with $T_{\rm eff}\gtrsim 8000\,\rm K$ rapidly lose  their narrow metal lines in the near infrared (Figure~\ref{fig:sensitivity}), making such companions very challenging to detect with APOGEE. One might still hope to detect the Brackett line cores of an RV-variable companion shifting from epoch to epoch, but this is not easily realized in practice because the emission line profiles of Be stars are themselves highly variable. 

As we show in Appendix~\ref{sec:sensitivity_appendix}, our search is thus sensitive to companions with $T_{\rm eff} \lesssim 8000\,\rm K$. Fortunately, the stripped companions in systems like LB-1 and HR 6819 are expected to have been cooler earlier in their evolution, when mass transfer was still ongoing. Emission lines are still expected to exist during this stage, but they will originate in an accretion disk, not in a decretion disk as in classical Be stars. 
This is the evolutionary stage to which the search is most sensitive.  

\begin{figure*}
    \centering
    \includegraphics[width=\textwidth]{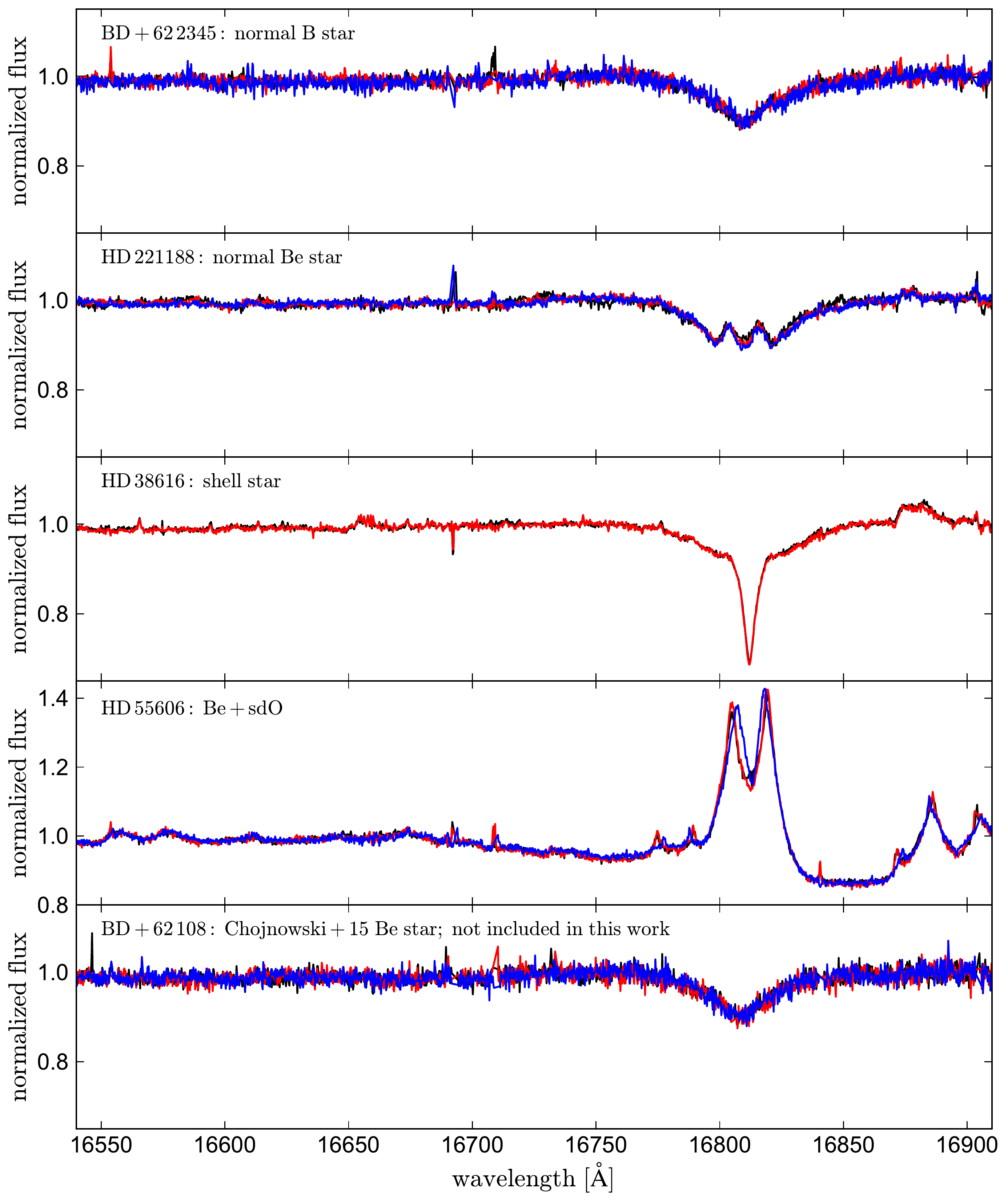}
    \caption{APOGEE spectral cutouts of representative B stars examined during our search. The strongest feature is the Br11 line at 16811\,\AA. Different colors show spectra from different visits. 1st and 2nd rows show examples of an normal B stars and Be stars, respectively. 3rd panel shows a shell star: a Be star viewed edge-on with a deep absorption line core owing to self-absorption by the disk. 4th panel shows the Be+sdO binary HD 55606 \citep[][]{Chojnowski2018}; note the RV shifts of several weaker emission lines. 5th panel shows an example of a star included in the \citet{Chojnowski2015} Be star sample but not in our sample. Any emission lines are subtle, if present at all. }
    \label{fig:example_spectra}
\end{figure*}

\subsubsection{Search results}

Our visual inspection resulted in three objects being flagged as potential Be star binaries. 

The first was HD 38616, whose spectra contain narrow lines but showed no evidence of RV variability between two visits separated by 3 days (3rd panel of Figure~\ref{fig:example_spectra}) . We obtained several high-resolution follow-up spectra with the FEROS spectrograph at La Silla \citep{Kaufer1999} over a 7 day period in December 2020. These confirmed the object to be a Be star with narrow absorption lines (spectral type B3) but ruled out RV variability of the absorption lines with amplitude larger than $1\,\rm km\,s^{-1}$ over the 7-day baseline. We conclude that this object is likely a normal ``shell star''; i.e., a Be star viewed nearly edge-on, with significant self-absorption from the circumstellar disk \citep[e.g.][]{Struve1931, Rivinius2006}. As one of the few Be stars in the TESS continuous viewing zone, it remains an interesting object to study in detail.

The 2nd object of interest we identified was HD 55606, whose spectra showed significant epoch-to-epoch line profiles variations in many emission lines and unusually strong metallic emission lines, but no obvious narrow absorption lines (4th panel of Figure~\ref{fig:example_spectra}). It turned out that this object was already studied in detail by \citet{Chojnowski2018} and \citet{Wang2021}, who found it to be a Be + sdO binary; i.e., a Be + stripped star binary in which the donor star has already contracted to become a hot subdwarf.

The final candidate was HD 15124, on which most of our analysis is focused. We describe the data for the system in Section~\ref{sec:data} and constrain  physical parameters in Section~\ref{sec:stellar_params}. Our constraints are summarized in Table~\ref{tab:system}.

\section{Data for HD 15124}
\label{sec:data}
\begin{table}
\centering
\caption{Physical parameters and $1\sigma$ uncertainties for both components of HD 15124.}
\begin{tabular}{lll}
\hline\hline
\multicolumn{3}{l}{\bf{Parameters of the unresolved source}}  \\ 
Right ascension & $\alpha$\,[deg] & $37.043254$ \\
Declination & $\delta$\,[deg] & $57.274803$ \\
Apparent {\it Gaia} eDR3 magnitude & $G$\,[mag] & $7.98$ \\
Apparent {\it Gaia} eDR3 color & $G_{\rm BP}-G_{\rm RP}$\,[mag] & $0.42$ \\

Corrected {\it Gaia} eDR3 parallax & $\varpi$\,[mas] & $1.64\pm 0.03$ \\
{\it Gaia} \texttt{ruwe} & & 1.11 \\

Color excess & $E(g-r)$\,[mag] & $0.255\pm 0.02$ \\
Dereddened absolute magnitude & $M_{G,0}$\,[mag] & $-1.62\pm 0.06$ \\

\hline

\multicolumn{3}{l}{\bf{Parameters of the subgiant donor star}}  \\ 
Effective temperature & $T_{\rm eff}$\,[K] & $6900 \pm 250$ \\
Surface gravity   & $\log(g/(\rm cm\,s^{-2}))$  & $2.8\pm 0.15$  \\
Projected rotation velocity & $v\sin i$\,[km\,s$^{-1}$] &  $20 \pm 3$ \\
Rotation velocity & $v_{\rm rot}$\,[km\,s$^{-1}$] &  $51 \pm 9$ \\
Synchronization parameter & $v_{{\rm rot}}/\left(2\pi R/P_{{\rm orb}}\right)$ &  $0.94 \pm 0.17$ \\
Microturbulent velocity & $v_{\rm mic}$\,[km\,s$^{-1}$] & $2.5\pm 1$ \\

Cont. flux ratio at 5000\,\textup{\AA} & $f_{{\rm donor}}/f_{{\rm tot}}$ & $0.20\pm0.02$ \\
Cont. flux ratio at 16000\,\textup{\AA} & $f_{{\rm donor}}/f_{{\rm tot}}$ & $0.35\pm0.03$ \\

Radius & $R\,[R_{\odot}]$ & $5.82 \pm 0.45$  \\ 
Bolometric luminosity & $\log(L/L_{\odot})$ & $1.83\pm 0.09$ \\ 
Mass &  $M\,[M_{\odot}]$ & $0.92\pm 0.22$ \\

\hline
\multicolumn{3}{l}{\bf{Parameters of the Be star}}   \\ 
Effective temperature & $T_{\rm eff}$\,[K] & $16000\pm 1000$ \\
Surface gravity   & $\log(g/(\rm cm\,s^{-2}))$  & $4.1 \pm 0.2$  \\
Projected rotation velocity & $v\sin i$\,[km\,s$^{-1}$] &  95 $\pm$ 10 \\
Cont. flux ratio at 5000\,\textup{\AA} & $f_{{\rm Be}}/f_{{\rm tot}}$ & $0.76\pm0.02$ \\
Cont. flux ratio at 16000\,\textup{\AA} & $f_{{\rm Be}}/f_{{\rm tot}}$ & $0.45\pm 0.05$ \\
Radius & $R\,[R_{\odot}]$ & $4.1 \pm 0.2$  \\ 
Mass &  $M\,[M_{\odot}]$ & $5.3 \pm 0.6$ \\ 
Bolometric luminosity & $\log(L/L_{\odot})$ & $3.00 \pm 0.12$ \\ 
Rotation velocity & $v_{\rm rot}$ & $241\pm 30$ \\ 
Fraction of critical rotation & $v_{\rm rot}/v_{\rm crit}$ & $0.60 \pm 0.08$ \\ 

\hline
\multicolumn{3}{l}{\bf{Parameters of the binary}}   \\ 
Orbital period & $P_{\rm orb}$\,[day]  & $ 5.4692 \pm 0.0001$  \\
Conjunction time & $T_0$\,[MHJD]  & $57678.740 \pm 0.001$  \\

Donor RV semi-amplitude  & $K_{\rm donor}$ [km\,s$^{-1}$] & $74.3\pm 0.9$ \\
Center-of-mass velocity & $\gamma_{\rm B}$\,[km\,s$^{-1}$] & $-11.1\pm 0.7$ \\ 
Donor eccentricity & $e_{\rm donor}$ & $<0.02$  \\
Donor RV scatter & $s_{\rm donor}$ [km\,s$^{-1}$] & $<1.5$ \\
Orbital inclination & $i\,[\rm deg]$ & $23.2\pm 1.5$  \\
Separation  & $a$ [R$_{\odot}$]   &  $24.0\pm 0.8$ \\
Mass function  & $f(M_{\rm Be})$ [M$_{\odot}$]   &  $0.23\pm 0.01$ \\

\hline

\multicolumn{3}{l}{\bf{Surface abundances -- assumed equal for both components }}  \\ 
Helium mass fraction & $Y_{\rm He}$  & $0.63 \pm 0.08$ \\
Carbon (relative to solar)  & C & $-2.3 \pm 0.2 $ \\
Nitrogen (relative to solar) & N  & $0.8 \pm 0.3 $ \\
Oxygen (relative to solar) & O & $-0.8 \pm 0.2 $ \\
Iron  (relative to solar) & Fe & $0.05 \pm 0.1$ \\

\hline
\multicolumn{3}{l}{\bf{Be star disk parameters}} \\
Cont. flux ratio at 5000\,\textup{\AA} & $f_{{\rm disk}}/f_{{\rm tot}}$ & $0.04\pm0.02$ \\
Cont. flux ratio at 16000\,\textup{\AA} & $f_{{\rm disk}}/f_{{\rm tot}}$ & $0.2\pm0.05$ \\
Emission line peak separation & $\Delta v_{\rm peak}$ [km\,s$^{-1}$] & $300\pm 50$ \\
Outer disk radius & $R_{\rm outer}\,[R_{\odot}]$ & $7.0 \pm 3.5$ \\

\hline
\end{tabular}
\begin{flushleft}

\label{tab:system}
\end{flushleft}
\end{table}

\subsection{Previous literature}
HD 15124 was classified as spectral type B5\,III ($T_{\rm eff} \approx 15000\,\rm K$) by \citet{Davis1973} and as B3/4\,V/IV ($T_{\rm eff} = 16000-18000\,\rm K$) by \citet{Jensen1981}. An $R\approx 3,000$ optical spectrum of the object was obtained by \citet{Huang2010}, who inferred $T_{{\rm eff}}=18603\pm300\,{\rm K}$ and $\log g = 4.21\pm 0.04$ by fitting the Balmer lines, and $v\sin i = 87\pm 13$ from the He I lines. The APOGEE ASPCAP pipeline found very different spectral parameters, $T_{\rm eff}=6200 \pm 65\,\rm K$ and $ \log g = 3.76 \pm 0.05$.  None of these analyses accounted for the presence of a second star or disk, so we suspect their reported uncertainties are significantly underestimated. \citet{Chojnowski2015} classified the object as a Be star based on its APOGEE spectra. To our knowledge, no previous studies recognized the object as a spectroscopic binary. 

\subsection{APOGEE spectra}
\label{sec:apogee}
\begin{figure*}
    \centering
    \includegraphics[width=\textwidth]{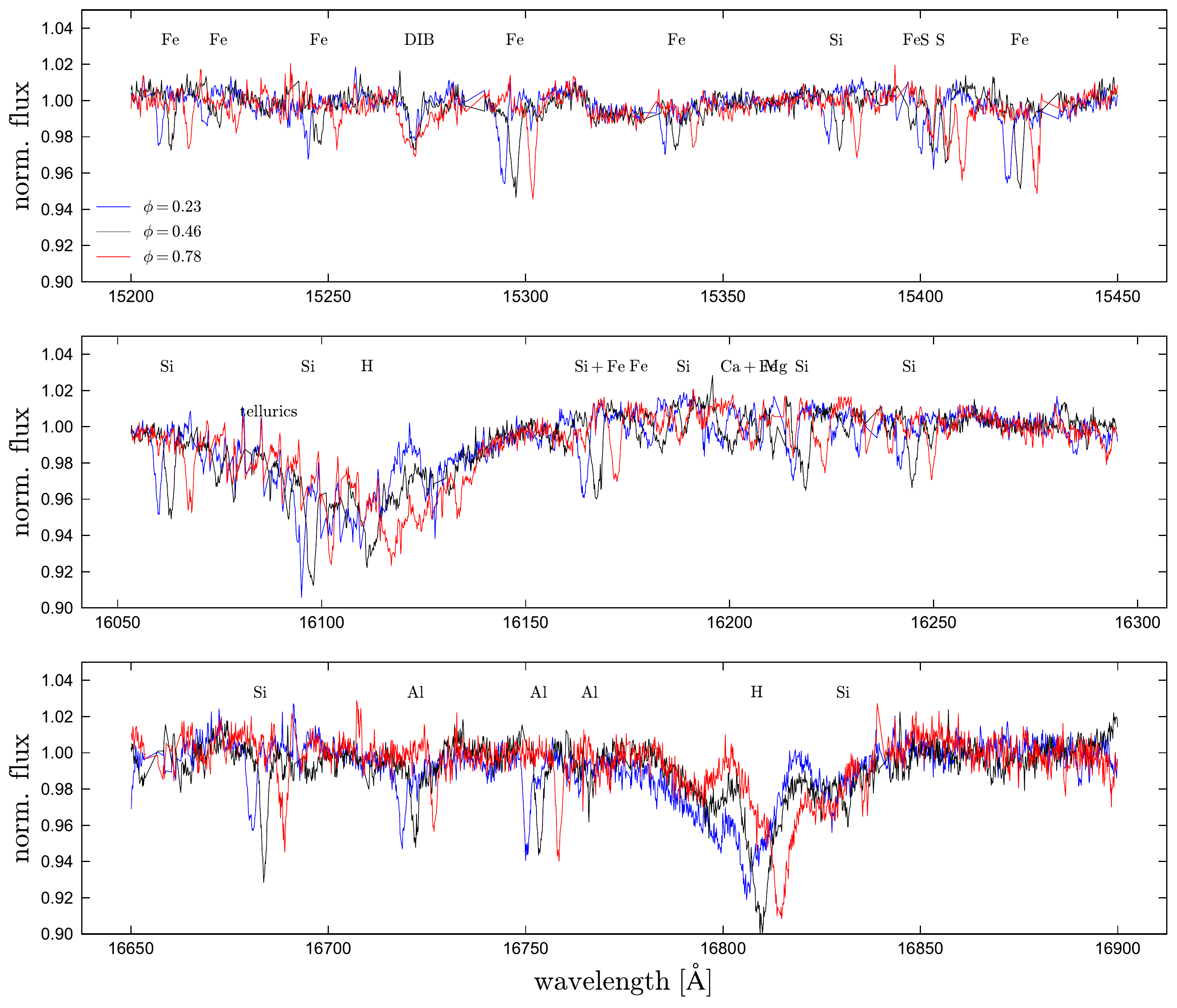}
    \caption{APOGEE spectra of HD 15124. We show spectra from three different visits, which are labeled by orbital phase (Section~\ref{sec:RVs}). All narrow absorption lines shift coherently, with a $\sim$140\,$\rm km\,s^{-1}$ shift between the visits plotted in blue and red. In addition to RV-variable absorption lines, double-peaked emission is evident in the hydrogen Brackett series (most obvious in Br11 at 16811\,\AA). RV shifts of the emission lines are weaker than those of absorption lines. Several strong lines are labeled; these all trace the narrow-lined, RV-variable donor star. }
    \label{fig:apogee_cutout}
\end{figure*}

HD 15124 was observed 11 times by the APOGEE survey between October 2016 and November 2017 (APOGEE ID \texttt{2M02281038+5716293}). The spectra have a typical SNR of 500 per pixel, wavelength coverage of 15150 to 16950 (with the standard two chip gaps), and spectral resolution $R\approx 22,500$. Figure~\ref{fig:apogee_cutout} shows three 250\,\AA-wide cutouts of the APOGEE spectra from three visits, which are chosen to span the range of RV shifts in the data. 

In contrast to the vast majority of Be stars, HD 15124 has narrow absorption lines, suggesting the presence of a slowly rotating star ($v\,\sin i\approx 20\,\rm km\,s^{-1}$; Section~\ref{sec:vsini}). Almost all the lines in the APOGEE spectra undergo coherent RV shifts from epoch to epoch, with a maximum shift of $\approx 145\,\rm km\,s^{-1}$. The exceptions are (a) diffuse interstellar bands (DIBs) at 15615, 15651 and 15671 \AA, and (b) double-peaked emission in the hydrogen Brackett lines (Br 11, 12, 13, 14, and 15 at 16811, 16411, 16114, 15885, and 15705\,\AA, with emission becoming weaker in the higher-order lines).  
As we will show, HD 15124 contains a cool, slowly-rotating star that is losing mass to a more-massive, hotter, rapidly rotating star with an accretion disk. We refer to the slowly-rotating, obviously RV-variable component as the ``donor'', and the accreting companion as the ``Be star''.

The flux calibration of APOGEE visit spectra is not expected to be precise \citep{Nidever2015}, so most of our analysis is based on pseudo-continuum normalized spectra. We define the pseudo-continuum using a spline fit to regions of the spectra without obvious absorption lines. This does not necessarily correspond to the ``true'' continuum (which is in any case ill-defined for a system containing two RV-variable stars and a disk), but simply allows us to bring observed spectra and models to a common scale. 

The narrow metal lines in the APOGEE spectra are relatively shallow, with the strongest lines having normalized depths ranging from 0.05 to 0.1. As we will show, this is in large part a consequence of continuum dilution by the Be star, which contributes more than half the flux in the $H$-band, but no narrow lines.

\subsection{Low-resolution, flux calibrated Shane/Kast spectrum}
\label{sec:kast}

\begin{figure*}
    \centering
    \includegraphics[width=\textwidth]{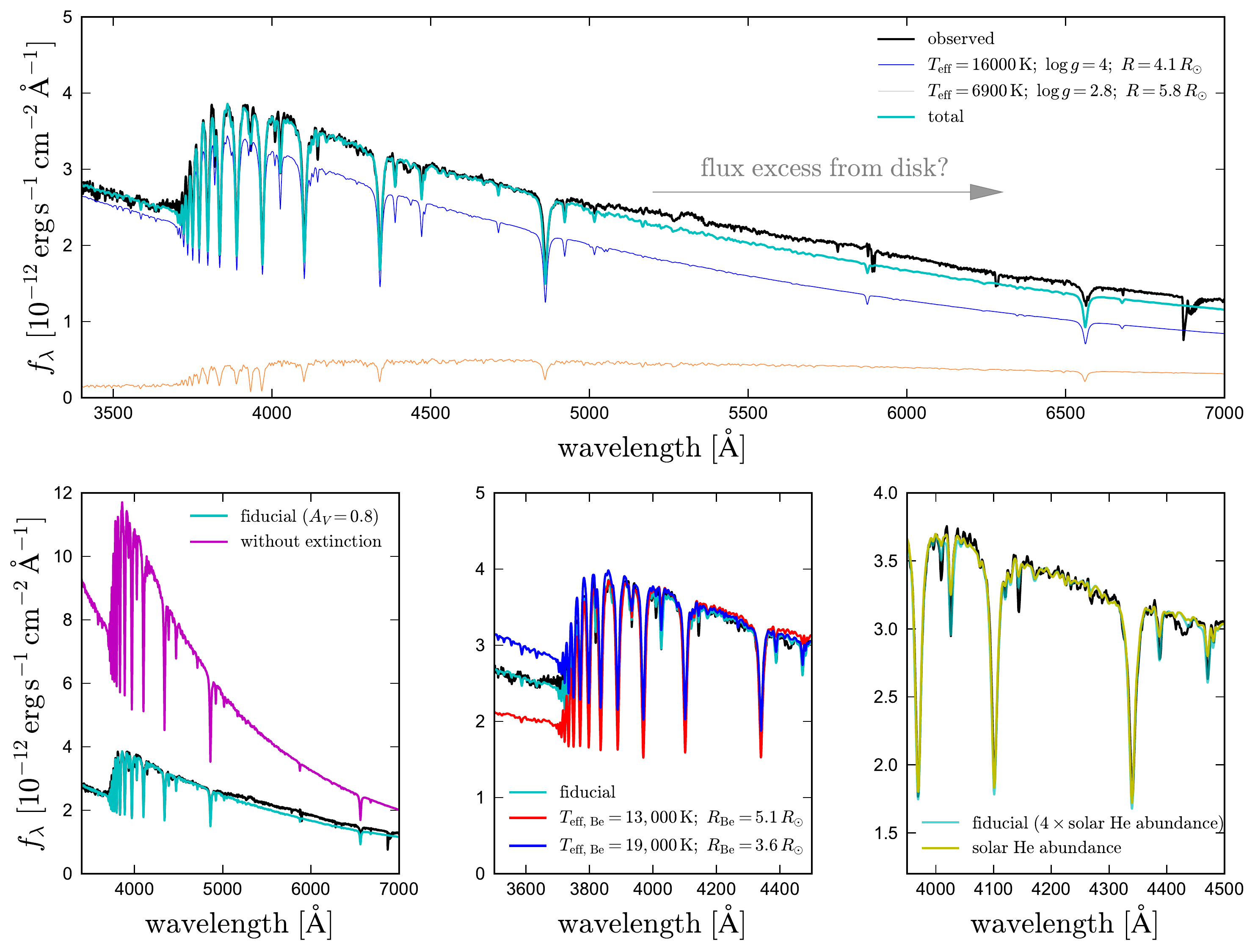}
    \caption{Flux-calibrated optical spectrum ($R\approx 1000$) of HD 15124 and spectral models. Top panel shows our best-fit model spectra, with the contributions of the Be star (blue) and donor (orange) shown separately. The Be star dominates the optical spectrum, with the donor contributing $\approx 20\%$. Bottom panels show the effects of extinction (left; we constrain $A_V$ based on the \citealt{Green2019} 3D dust map), the temperature of the Be star (middle; it is constrained by the strength of the Balmer jump, and the Be star radius is adjusted to match the normalization at 4000\,\AA\,\, for each temperature), and helium abundance (right; the strong observed He I lines require significant He enhancement).} 
    \label{fig:fluxed}
\end{figure*}

We observed HD 15124 using the Kast spectrograph \citep{Miller1994} on the 3m Shane telescope at Lick Observatory on November 3, 2020.  We used the 600/7500 grating on the red side and the 600/4310 grism on the blue side, with the D55 dichroic and a 1 arcsec slit. This resulted in wavelength coverage of 3300–8400 \AA\,\,with spectral resolution $R\approx 1,000$. We observed the A3V standard star HD 37725 for flux calibration. A 120-second exposure yielded SNR $\approx 300$ per pixel at 4000\,\AA. We reduced the spectrum using \texttt{pypeit} \citep{Prochaska_2020}.

The spectrum is shown in Figure~\ref{fig:fluxed}.
Most narrow metal lines from the donor are not detectable in the spectrum due to the low spectral resolution; most of the obvious features come from the Be star. A notable exception is the Ca II H\,\&\,K lines, which are strong in the donor but weak in the hotter Be star. 
Although its spectral resolution is low, the Kast spectrum provides strong constraints on the temperature and radius of the Be star. In particular, the Balmer jump at 3646 \AA\ is sensitive primarily to the Be star's effective temperature, with weak dependence on its surface gravity and composition. With only a low-resolution spectrum, it would be difficult to determine how much of the optical flux comes from the donor. Fortunately, the donor's contributions to the spectrum are constrained reliably from the high-resolution spectrum.

\subsection{High-resolution LBT/PEPSI spectrum}
\label{sec:pepsi}

\begin{figure*}
    \centering
    \includegraphics[width=\textwidth]{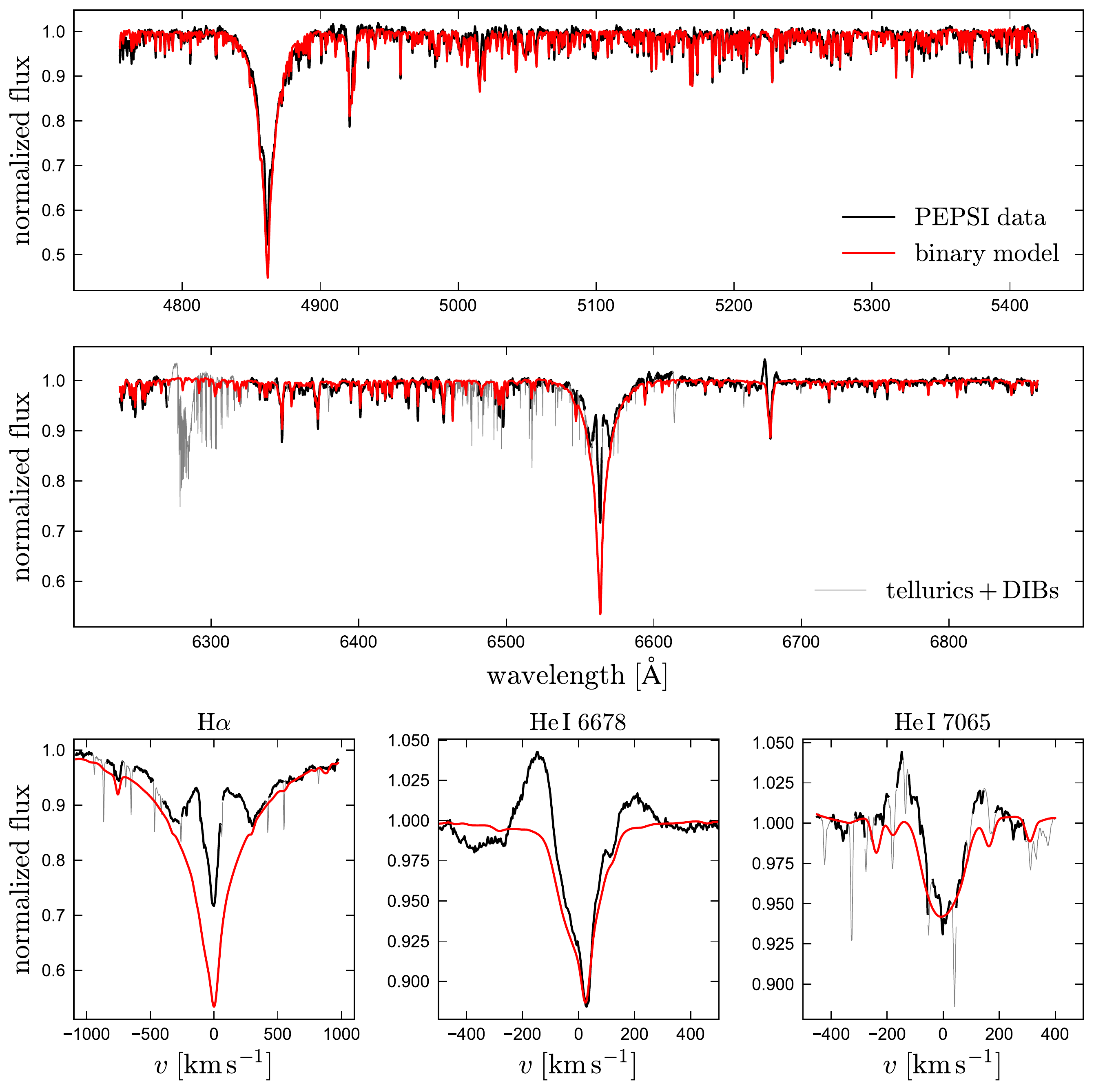}
    \caption{PEPSI spectrum of HD15124 (black) and best-fit binary model (red; see Section~\ref{sec:stellar_params} for details). Greyed-out regions of the spectra are masked due to contamination from tellurics and interstellar absorption. Bottom three panels show regions of the spectra with prominent emission lines from the disk. Most of the visually obvious absorption lines are from the cool donor star (orange line in the upper panel of Figure~\ref{fig:fluxed}), which contributes only $\sim 20\%$ of the light. The contributions of the Be star are most evident in the Balmer and He I lines.  }
    \label{fig:pepsi}
\end{figure*}

We observed HD 15124 for 300 seconds on September 28, 2020 using the Potsdam Echelle Polarimetric and Spectroscopic Instrument (PEPSI; \citealt{Strassmeier2015}) spectrograph on the Large Binocular Telescope in binocular mode. The spectrum was reduced as described in \citet{Strassmeier2018}; it covers the wavelength ranges of 4755-5428\,\AA\,\,and 6236-7433\,\AA\,\,with spectral resolution $R\approx 50,000$ and per-pixel SNR\,$\approx$\,500. The spectrum was obtained at phase $\phi= 0.59$ (see Section~\ref{sec:RVs}). Figure~\ref{fig:pepsi} shows portions of the spectrum and our best-fit model (Section~\ref{sec:stellar_params}). Zoomed-in cutouts of the spectrum are also shown in Figures~\ref{fig:vturb} and~\ref{fig:abundances}. 

Like the APOGEE spectra, the PEPSI spectrum contains double-peaked emission lines. Emission is obvious in H$\alpha$ and more subtle, but still unambiguously present, in H$\beta$. There is also double-peaked emission in some, but not all, of the He I lines, with a typical separation of $\approx 300\,\rm km\,s^{-1}$ between the peaks. There are no detectable Fe lines in emission. 

The PEPSI spectrum makes it obvious that at least two luminous stars contribute in the optical, because it contains features characteristic of both hot and cool stars. In particular, the presence of many narrow metal lines (all due to the donor) suggests a temperature near 7000\,K, while the presence of moderately strong helium line in absorption (due to the Be star) suggests a temperature of at least 15,000\,K. Fitting the PEPSI and Kast spectra simultaneously (Section~\ref{sec:stellar_params}) allows us to constrain the atmospheric parameters and abundances of both components.


\subsection{NRES spectra}
\label{sec:lco}

We observed HD 15124 with the NRES spectrograph \citep{Siverd2018} on the 1\,m telescope at Wise observatory through the Las Cumbres Observatory Global Telescope network \citep{Brown2013} 32 times between February 1 and March 8 of 2021. The wavelength coverage is 3800-8200\,\AA, with spectral resolution $R\sim 30,000$. Exposure times ranged from 600 to 1800 seconds, yielding a typical per-pixel SNR of 5 at 5,000\,\AA. The spectra were reduced with the Banzai-NRES pipeline\footnote{https://github.com/LCOGT/banzai-nres}. The low SNR of these spectra made them useful primarily for measuring RVs, which we used to constrain the binary's orbit (Section~\ref{sec:RVs} and Figure~\ref{fig:rvs}).

\subsection{Light curves: Hipparcos, WISE, TESS, and KELT}
\label{sec:lightcurves}

To search for photometric variability, we retrieved light curves of HD 15124 from several time-domain photometric surveys. Unfortunately, the star is so bright that its photometry is saturated in many time-domain surveys (e.g. ASAS-SN, ZTF, CSS). However, the available photometry still allows us to understand the object's photometric variability over a range of timescales and wavelengths. 

\begin{figure*}
    \centering
    \includegraphics[width=\textwidth]{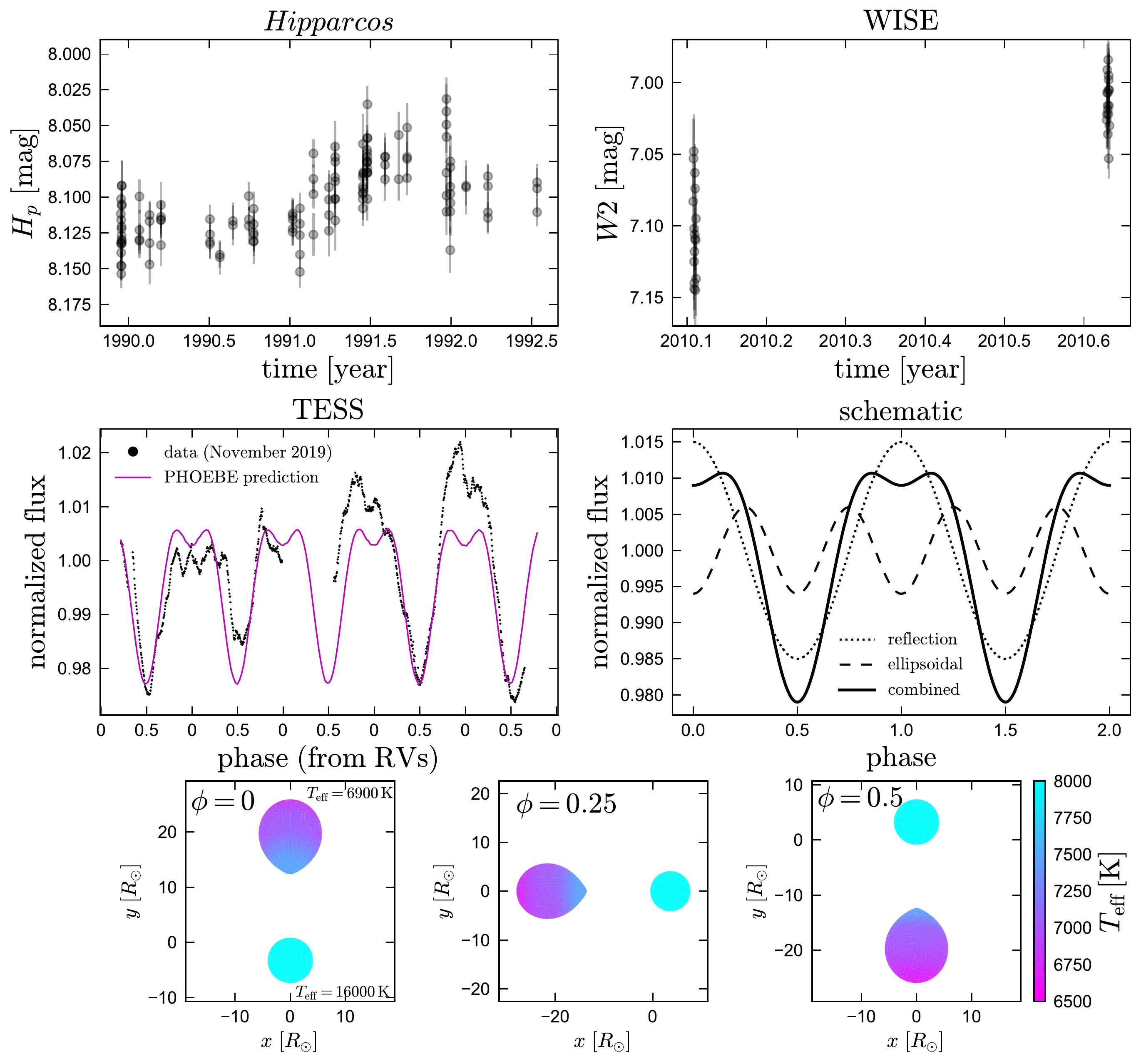}
    \caption{Light curves for HD 15124. Top panels show data from {\it Hipparcos} (over 3 years) and WISE (two sets of observations separated by half a year). On long timescales, the source's brightness varies at the 0.05-0.1 mag level, likely due to changes in the structure of the Be star's accretion disk. Center left panel shows the TESS light curve, which spans 25 days. The data is phased based on the RV ephemeris (Section~\ref{sec:RVs}). We overplot a model light curve calculated with PHOEBE for the stellar parameters calculated in Section~\ref{sec:stellar_params} (not a fit). The variability in the PHOEBE light curve is primarily due to reflection and ellipsoidal variation, as illustrated schematically in the center right panel. Bottom panels show the mesh corresponding to the PHOEBE light curve at three different phases. Reflection (i.e., irradiation of one side of the donor by the Be star) dominates the short-timescale variability.}
    \label{fig:light_curves}
\end{figure*}

\begin{figure*}
    \centering
    \includegraphics[width=\textwidth]{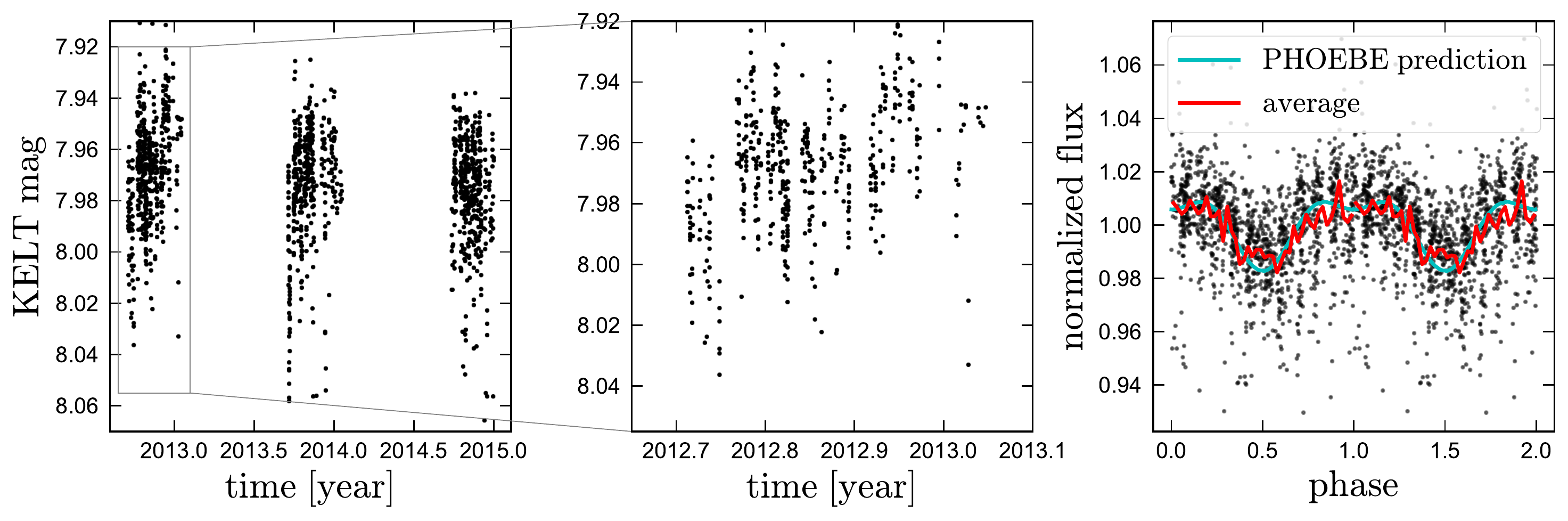}
    \caption{KELT optical light curve of HD 15124. Left panel shows the raw light curve, which contains $\sim$1300 measurements over 2.5 years. A few points with much fainter magnitudes, likely due to systematics, fall outside the axis limits. Middle panel shows a zoom-in on a 5-month period (30 orbits), revealing variability on a range of timescales. Right panel shows the full light curve, phased to the ephemeris from the RVs. The phased light curve has similar shape to the TESS light curve (Figure~\ref{fig:light_curves}), revealing a combination of ellipsoidal variability and the reflection effect. }
    \label{fig:kelt}
\end{figure*}

\subsubsection{Hipparcos}
The {\it Hipparcos} mission \citep{ESA1997} produced 132  usable photometric measurements of HD 15124 (HIP 11487) over a 3 year baseline with median uncertainty of 0.013 mag. We retrieved the light curve from the ESA archive (Disk 2, Volume 17) and show it in the upper left panel of Figure~\ref{fig:light_curves}. The sparse sampling of the data make it difficult to probe short-timescale variability, but longer-timescale variability is obvious. Most notably, the source brightened by about 0.05 mag in the first half of 1991, and then faded again by $\sim$0.02 mag in 1992. 

\subsubsection{WISE}
HD 15124 was also observed by the WISE mission \citep{Wright2010}. We retrieved multi-epoch photometry from the AllWISE Multiepoch Photometry database \citep[][]{Cutri2021}, which includes usable light curves in both the W1 and W2 bands. The W2-band light curve, which has typical uncertainty 0.022 mag, is shown in the upper right panel of Figure~\ref{fig:light_curves}. The behavior of the W1 light curve is similar, but its uncertainties are larger. Like the {\it Hipparcos} light curve, the WISE data shows clear evidence of long-term flux variability, with the mean magnitude brightening by 0.09 mag over a half-year period. 

Long-timescale brightness variability in both the optical and infrared is ubiquitous in Be stars \citep[e.g.][]{Rivinius2013}. In most cases, variability is attributed to growth and shrinking of a circumstellar disk. Variability is typically more pronounced in the infrared, because the disk contributes a larger fraction of the total light there. The amplitude of the optical and NIR variability observed for HD 15124 ($\lesssim 0.1$ mag) is on the low side for classical Be stars, where variability amplitudes of $\approx 0.5$ mag are common \citep[e.g.][]{Labadie-Bartz2017}. The accretion disk in HD 15124 -- being truncated by a close companion -- must be smaller than the disks in classical Be stars, which often extend to tens of stellar radii. 

\subsubsection{TESS}
HD 15124 was observed by the TESS satellite \citep[][]{Ricker2015} for 25 days during sector 18 in November 2019. We used \texttt{eleanor} \citep{Feinstein2019} to perform background subtraction and PSF photometry on the TESS full-frame images, extracting a light curve with 30 minute cadence. We experimented with a variety of apertures, finding the shape of the extracted light curve relatively robust as long as the central pixels nearest to HD 15124 are included. The star has no comparably bright neighbors within several arcminutes, so contamination from background sources is expected to be modest. 

The extracted light curve, which spans 25 days and includes a 3-day gap for data downlinking, is shown in the middle left panel of Figure~\ref{fig:light_curves}. We calculate the phase of the TESS observations using the orbital ephemeris from the RVs (Section~\ref{sec:RVs}), which is defined such that the Be star would eclipse the donor (if the inclination were close to edge-on) at phase 0. The TESS light curve reveals complex variability on short timescales. Most pronounced is quasi-sinusoidal variability on the orbital period, with semi-amplitude of (1-2)\%. Additionally, there is short-timescale (hours to days) variability that is manifest as jagged, somewhat irregular fluctuations in the light curve, as well as longer-timescale variations. We do not find comparable variability in the TESS light curves of most other nearby bright stars, suggesting that both the short- and long-timescale variability observed in HD 15124 is astrophysical. Irregular photometric variability is common in accretion disks due to a combination of turbulence and changes in disk structure \citep[e.g.][]{Bruch_1992}. 

\subsubsection{KELT}
\label{sec:kelt_lc}
HD 15124 was observed by the Kilodegree Extremely Little Telescope  \citep[KELT;][]{Pepper2007} between 2012 and 2015, yielding 1355 150-second exposures. The data were obtained using a 42 mm-aperture telescope at Winer Observatory with a $26 \times 26$ degree field of view and a broad photometric filter that transmits most light redward of 5000\,\AA. Along with raw light curves, the KELT pipeline produces a detrended light curve, which filters out longer-timescale variability and systematics common with other nearby stars using the Trend Filtering Algorithm \citep[TFA;][]{Kovacs2005}. HD 15124 falls in KELT field N17, the photometry of which is as yet unpublished. 

The KELT light curve is shown in Figure~\ref{fig:kelt}. To approximately convert the KELT instrumental magnitudes to Johnson-Cousins {\it R}-band, we subtract 4.1 from the reported values. Like the TESS data, the raw light curve (left and middle panels) reveals both short- and long-timescale variability. To reduce scatter due long-term variability, we crudely detrended the raw light curve using a running median filter with width of 33 days (6 orbital periods).\footnote{We also experimented with using the TFA detrended light curve. This yielded similar results, with somewhat less scatter, but also with a $\sim$20\% lower variability amplitude, suggesting that the detrending removed some real astrophysical variability.} We also clipped outliers with calibrated KELT magnitudes outside the range (7.90, 8.05). The resulting phase-folded light curve (right panel) has a similar shape to the TESS light curve, suggesting that the light curve shape is relatively stable over long timescales. 

\subsubsection{Origin of the photometric variability}
\label{sec:light_curve_model}
Also plotted in Figure~\ref{fig:light_curves} is a model light curve computed with PHOEBE \citep{Prsa2005, prsa2016, horvat2018}, a program which builds on the \citet{Wilson1971} code for modeling binary light curves. The model we show is not a fit to the observed light curve, but is a prediction based on the binary parameters estimated from the spectra and RVs (Section~\ref{sec:stellar_params}). Although it does not reproduce all the observed variability, the model does explain the most prominent feature in the light curve: variability on the orbital period, with minima that are sharper than maxima, almost resembling eclipses. 

This light curve shape results from the combined effects of reflection/irradiation and tidal distortion of the donor. The irradiated, Be star-facing side of the donor is hotter and brighter than the opposite side. This alone leads to approximately sinusoidal variability on the orbital period. In addition, the donor star is tidally distorted into a ``teardrop'' shape, such that the geometric cross-section visible to an observer changes over the course of the orbit. This also produces nearly sinusoidal variability, but with a period of half the orbital period. Adding these two effects together leads to the light curve shape predicted by PHOEBE, with sharp minima and flat maxima (middle right panel of Figure~\ref{fig:light_curves}).\footnote{The light curve morphology -- both the periodic and irregular behavior -- are similar to that of the mass-transfer binary HD 63021 recently studied by \citet{Whelan2021}. We suspect that the origin of the photometric variability in that system, which Whelan et al. did not explain, is also a combination of reflection and ellipsoidal variability.} The origin of this variability is illustrated further in the bottom panels of Figure~\ref{fig:light_curves}, which show the scene used by PHOEBE to calculate the model light curve at different phases. Due to reflection and ellipsoidal distortion, the surface temperature of the donor star is predicted to vary significantly (by about 1000\,K) over the star's surface, with the hottest point being the side facing the Be star. This means that the flux-weighted mean temperature (i.e., what we measure spectroscopically) is expected to vary somewhat with orbital phase. This is indeed what we find from the APOGEE spectra: the spectroscopic effective temperature is $\approx 500$\,K hotter at phase 0 than at phase 0.5. 

As we discuss in Section~\ref{sec:inc}, the inclination of HD 15124 is close to face-on ($i \sim 23\%$). The amplitude of variability due to reflection and ellipsoidal variability scale respectively as $\sin i$ and $\sin^2 i$ \citep[e.g.][]{Faigler2011}, so the observed variability is significantly weaker than would be expected for a similar binary viewed close to edge-on. The additional light from the Be star, which is not significantly irradiated or tidally distorted, further dilutes the variability amplitude.

\subsection{Radial velocities}
\label{sec:RVs}

We measured radial velocities of the donor in HD 15124 by cross-correlating a model spectrum with both the APOGEE spectra and the NRES spectra. We only consider the NRES spectra from which we obtained RV uncertainties of less than 20\,$\rm km\,s^{-1}$, as the rest were obtained in poor conditions and are consistent with noise. This left us with 11 RVs from APOGEE and 29 from NRES, spanning a 4 year baseline. We were unable to measure RVs for the Be star, because its contributions to the APOGEE spectra are subtle (primarily excess emission in the Brackett series), and the NRES spectra do not have sufficient SNR to extract reliable RVs from the helium absorption lines. 

\begin{figure*}
    \centering
    \includegraphics[width=\textwidth]{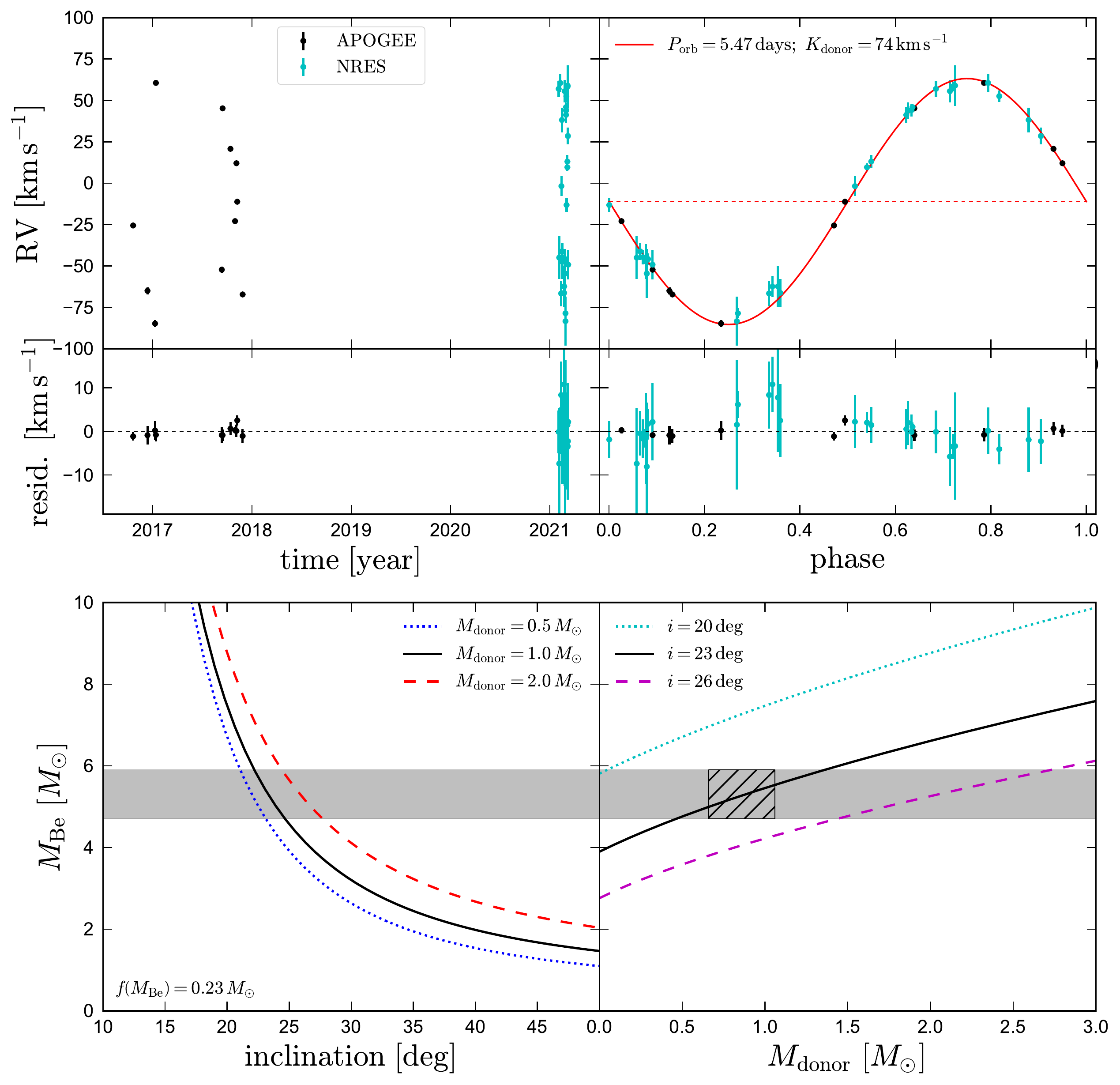}
    \caption{Upper panels: radial velocities of the donor. Black and cyan point show RVs measured from APOGEE spectra and follow-up observations with NRES. Right panel shows phased RVs and the corresponding orbital solution. Bottom panels show the dynamically-implied mass of the Be star, given the observed mass function, $f\left(M_{{\rm Be}}\right)=0.23\,M_{\odot}$, for a range of donor masses and inclinations. Gray shaded region shows plausible Be star masses, given the observed temperature and radius (Figure~\ref{fig:be_star_teff_logg}). Together, these constrain the orbital inclination to $i\approx 23.2\pm 1.5$ degrees. Hatched region shows the donor mass constraint if the donor fills its Roche lobe (Figure~\ref{fig:roche}). }
    \label{fig:rvs}
\end{figure*}

The resulting RVs are listed in Table~\ref{tab:rvs} and shown in the upper panels of Figure~\ref{fig:rvs}. We fit them with a Keplerian orbit using a combined simulated annealing + Markov chain Monte Carlo method, as described in \citet{El-Badry2018}. We also fit an RV ``scatter'' term $\sigma_{\rm scatter}$, which we add in quadrature to the measured RV uncertainties to represent either underestimated uncertainties or intrinsic scatter in the RVs.  Given the large number of RVs, the posterior is reasonably well-behaved and the period is unambiguously determined to be $P_{\rm orb}\approx 5.47$ days. We first left the eccentricity free and found $e\approx 0.02\pm 0.01$; i.e., marginal evidence of eccentricity. Due to the short orbital period and tidal deformation of the donor, we expect the orbit to be tidally circularized and suspect that the nonzero eccentricity favored by the initial fit is simply a manifestation of RV scatter. We therefore fixed the eccentricity to 0, leading to an inferred $\sigma_{\rm scatter}\approx 1\,\rm km\,s^{-1}$. Nonzero scatter (with a larger magnitude) was also found in HR 6819 \citep{Bodensteiner2020, El-Badry2021}, where it is  likely due to non-radial pulsations in the donor. The scatter in HD15124 might have similar origin, but might also stem from long-term RV variability due a tertiary (Appendix~\ref{sec:disc_triple}), or differences between the NRES and APOGEE RV zeropoints.

The orbital solution of the donor yields a joint constraint on the component masses and inclination, as described in Section~\ref{sec:inc}.

\subsection{Broadband SED}
\label{sec:SED}
\begin{figure*}
    \centering
    \includegraphics[width=\textwidth]{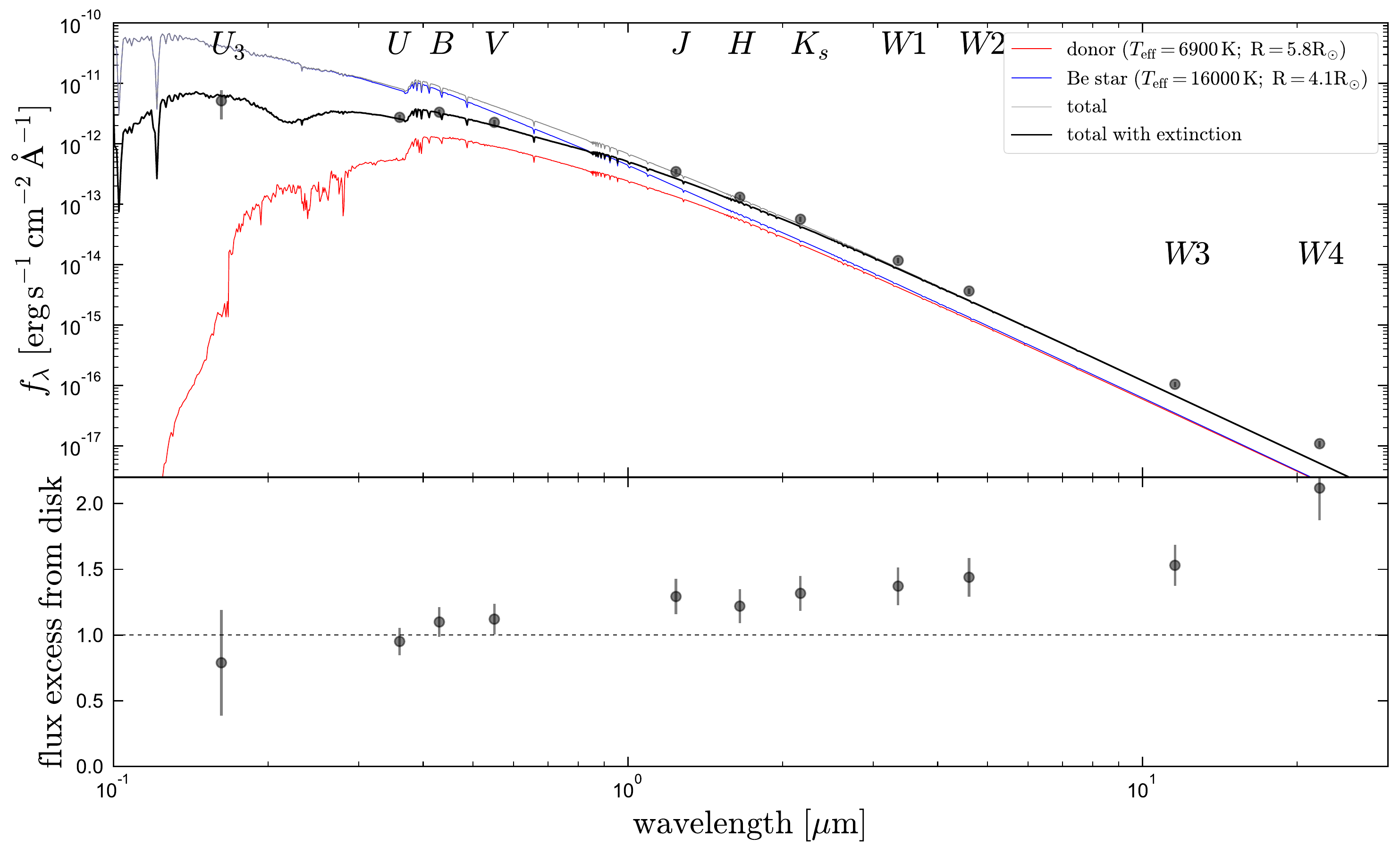}
    \caption{Broadband spectral energy  distribution. Points with error bars show measurements from the Celescope UV mission (``$U_3$''; \citealt{Davis1973}), ground-based UBV photometry \citep{Deutschman1976}, 2MASS ($JHK_s$; \citealt{Skrutskie2006}), and WISE ($W1W2W3W4$; \citealt{Wright2010}). Blue and red lines show model  spectra for the Be star and donor; black line shows their sum after accounting for extinction ($A_V=0.8$ mag). These are not a fit to the full observed SED, but are scaled to match the blue optical spectrum (Figure~\ref{fig:fluxed}). Bottom panel shows the ratio of the observed fluxes to the predicted bandpass-integrated magnitudes. There is definite evidence of IR excess from the disk, which contributes $\sim 20\%$ of the total light in the $H$ band, and more at longer wavelengths.  }
    \label{fig:sed}
\end{figure*}

To estimate the contributions of the circumstellar disk as a function of wavelength, we compared the source's measured fluxes from broadband photometry to synthetic photometry predicted based on the parameters of the donor and Be star alone. Johnson-Cousins $UBV$ photometry for HD 15124 was measured from the ground by \citet{Deutschman1976}, while near-IR and IR photometry was obtained by the 2MASS \citep{Skrutskie2006} and WISE \citep{Wright2010} missions. No UV observations of HD 15124 are published from GALEX \citep{Morrissey2007}. Fortuitously, a UV flux was measured by one of the earliest UV missions, the Celescope experiment \citep{Davis1968}. We converted the measured Celescope magnitude, $U_3 = 11.53$ mag \citep{Davis1973} to a flux density using the empirical calibration from \citet{Houziaux1974}. This yielded a well-sampled SED from 0.1 to 20 $\mu$m, which is plotted in Figure~\ref{fig:sed}. We inflate all the reported photometric uncertainties by adding a 10\% flux uncertainty in quadrature to the reported values to conservatively account for HD 15124's photometric variability and the unknown phasing of the photometry.

We predict bandpass-averaged mean magnitudes for both components using empirically-calibrated theoretical models from the BaSeL library \citep[v2.2;][]{Lejeune1997, Lejeune1998}. We assume a \citet{Cardelli_1989} extinction law with total-to-selective extinction ratio $R_V =3.1$, and we use an extinction prior from the \citet{Green2019} 3D dust map. We use \texttt{pystellibs}\footnote{\href{http://mfouesneau.github.io/docs/pystellibs/}{http://mfouesneau.github.io/docs/pystellibs/}} to interpolate between model spectra, and \texttt{pyphot}\footnote{\href{https://mfouesneau.github.io/docs/pyphot/}{https://mfouesneau.github.io/docs/pyphot/}}  to calculate synthetic photometry. In the bottom panel of Figure~\ref{fig:sed}, we plot the ratio of the observed fluxes and those predicted from the model of the donor and Be star, neglecting the disk. The observed UV and UBV fluxes are consistent with coming only from the Be star and donor. This is almost by construction, since we required the model spectra to match the flux calibrated blue-optical spectrum (Figure~\ref{fig:fluxed}). However, there is clear evidence of excess emission at NIR and IR wavelengths, as is very common for Be stars.

\subsection{Astrometry}
\label{sec:astrometry}
After correcting for the parallax zeropoint \citep{Lindegren2021} and underestimated parallax uncertainties for bright stars \citep{El-Badry2021_wb}, the {\it Gaia} eDR3 parallax of HD 15124 is $\varpi = 1.64\pm0.03$ (eDR3 source id 458048851856088064; \citealt{Gaia2021}). This corresponds to a distance $d = (610\,\pm 11)\,{\rm pc}$. The reported \texttt{ruwe} value is 1.11, indicating a reasonably good single-star astrometric solution. For our best-fit parameters (Section~\ref{sec:stellar_params}), the projected semimajor axis of the binary is $\rho=\varpi\times\left(a/1\,{\rm au}\right)\approx0.17\,{\rm mas}$, about 11\% of the parallax. This may account for the somewhat larger than average \texttt{ruwe} \citep[e.g.][]{Belokurov2020} but is small enough that we expect the parallax to be reliable. 

Although the {\it Gaia} astrometric solution appears to be well-behaved, comparison of the {\it Hipparcos--Gaia} position change with the measured {\it Gaia} proper motion reveals evidence of long-term acceleration of the photocenter \citep{Kervella2021}. We discuss this in detail in Appendix~\ref{sec:disc_triple}, where we conclude that HD 15124 may be a triple with a faint tertiary. The expected mass of the putative tertiary ($\sim 1\,M_{\odot}$) is low enough that it would be expected to contribute less than one percent of the total luminosity, so it has little effect on our modeling of the system. 

\section{Parameters of the binary}
\label{sec:stellar_params}

To constrain the atmospheric parameters, radii, and abundances of both stars, as well as the distance and extinction, we simultaneously fit the flux-calibrated low-resolution spectrum (Figure~\ref{fig:fluxed}) and the normalized high-resolution spectra (Figures~\ref{fig:apogee_cutout} and~\ref{fig:pepsi}). The flux-calibrated spectrum primarily constrains the temperature and radius of the Be star, while the high-resolution spectrum constrains the donor atmospheric parameters, radii, projected rotation velocities, and abundances of both stars.

\subsection{Spectral fitting and uncertainties}
\label{sec:uncert}

We simultaneously fit the Kast, PEPSI, and APOGEE spectra. We generated an irregular grid of  1D LTE atmospheres and synthetic spectra using ATLAS 12 \citep{Kurucz_1970, Kurucz_1979, Kurucz_1992} to compute the atmosphere structure and SYNTHE \citep{Kurucz_1993} for the radiative transfer calculations. We re-computed the atmosphere structure for each combination of atmospheric parameters and abundances. We use the linelist maintained by R. Kurucz.\footnote{http://kurucz.harvard.edu/linelists.html} To interpolate between models and predict spectra with atmospheric parameters and abundances not represented in the  grid, we used a 2nd-order polynomial spectral model \citep[e.g.][]{Rix2016, Ting2019}. Models were generated at resolution $R=300,000$, broadened using the rotation kernel from \citet{Gray_1992}, and convolved with a Gaussian line spread function with spectral resolution (FWHM) $R=50,000$ for PEPSI spectra, $R=22,500$ for APOGEE spectra, and $R=1,000$ for Kast spectra. 

We pseudo-continuum normalize the APOGEE and PEPSI spectra using a 3rd order spline fit to regions without strong absorption lines, and we normalize the model spectra in the same way. When predicting model normalized spectra for the unresolved binary, we add unnormalized model fluxes for the two stars prior to normalization, as described in \citet{El-Badry2018_theory}. For the model APOGEE spectra, we include 20\% continuum dilution from the Be star's disk, as suggested by the SED (Figure~\ref{fig:sed}). We do not normalize the Kast spectrum, which sets the flux scale, and thus the radii of both stars.

We began with a spectral model with 10 labels: the effective temperatures, surface gravities, radii, and projected rotation velocities of both stars, the microturbulent velocity $v_{\rm turb}$ of the donor, and a global metallicty [Fe/H], varying all metals in lockstep with [Fe/H] and assuming the solar abundance pattern. It quickly became clear that this was not a sufficient set of abundance labels, because (a) the model helium lines were weaker than observed, and (b) carbon lines in the observed spectra are extremely weak, while other metal lines are relatively strong. This suggested that the abundances are affected by CNO processing. We therefore added the helium, carbon, nitrogen, and oxygen abundances as additional labels to be fit. 

We implicitly assume that the two stars have the same surface abundances. In practice, the metallicity and CNO abundances are well-constrained only in the donor, since metal lines are rotationally smeared out in the Be star, and the He abundance is constrained only in the Be star, due to the donor's cooler temperature. We mask the Balmer and Brackett lines during fitting, since they are contaminated by emission from the disk. We also mask the two helium lines with obvious emission, and regions that are affected by tellurics or interstellar absorption (Figure~\ref{fig:pepsi}).

\begin{figure*}
    \centering
    \includegraphics[width=\textwidth]{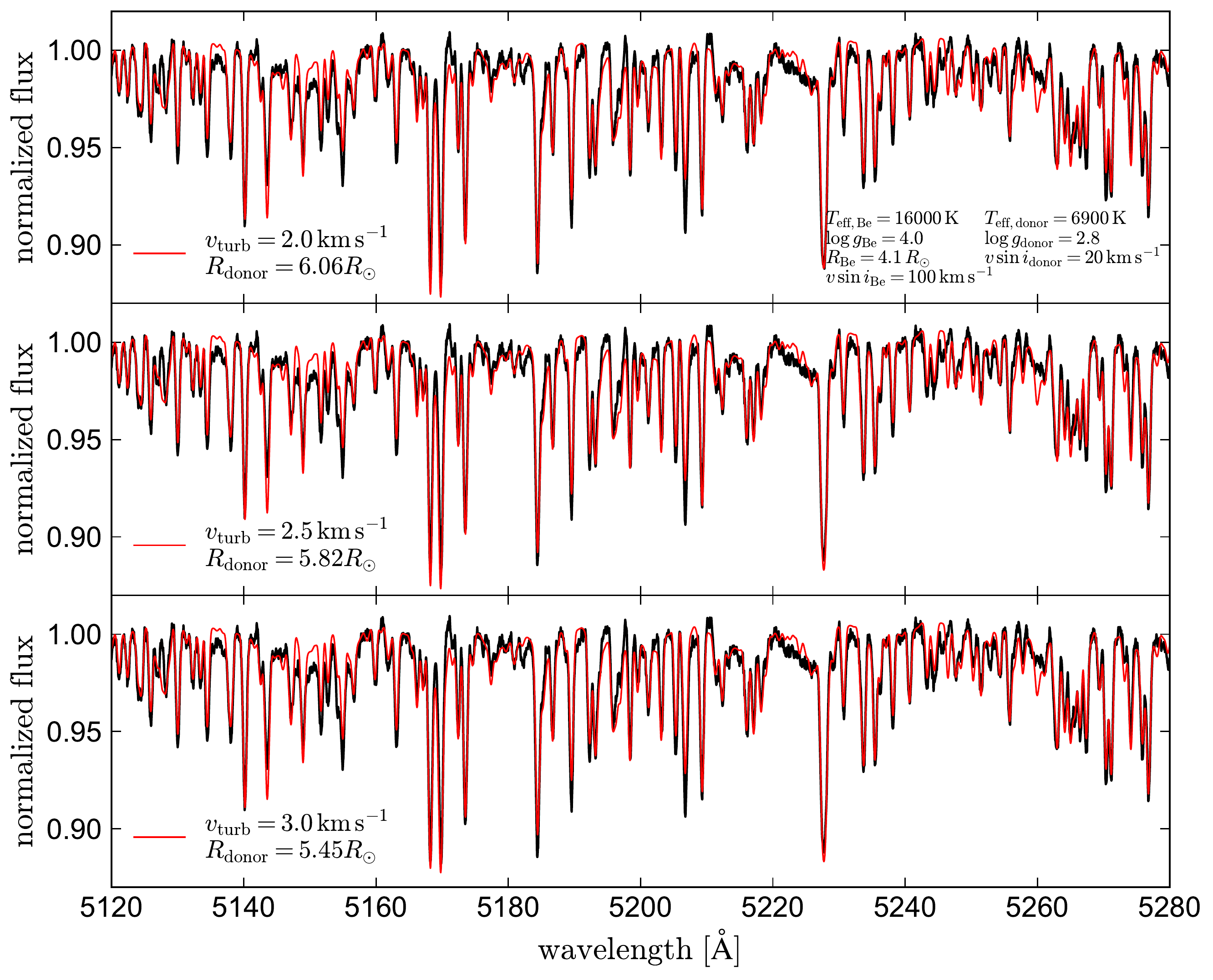}
    \caption{PEPSI spectrum and best-fit Kurucz model spectra for three different values of $v_{\rm turb}$, the microturbulent velocity. In all panels, we fix the parameters of both components to the values listed in the top panel. We fix $v_{\rm turb}$ to a different value in each panel and solve for the $R_{\rm donor}$ value that best fits the observations given this  $v_{\rm turb}$. Increasing $R_{\rm donor}$ increases the fraction of the flux contributed by the donor. Conversely, increasing $v_{\rm turb}$ makes most of the donor's lines deeper. There is thus a significant covariance between $v_{\rm turb}$ and $R_{\rm donor}$. For the best-fit $v_{\rm turb}=2.5\,\rm km\,s^{-1}$ and $R_{\rm donor}=5.8\,R_{\odot}$, the donor contributes 21\% of the total flux at 5200\,\AA. }
    \label{fig:vturb}
\end{figure*}

Among the largest sources of uncertainty in fitting the spectrum is the micrturbulence of the donor, which is covariant with the flux ratio (and thus with $R_{\rm donor}$ and $T_{\rm eff,\,donor}$). This is illustrated in Figure~\ref{fig:vturb}, which shows best-fit models for the PEPSI spectrum for three different values of $v_{\rm turb}$. Increasing $v_{\rm turb}$ increases the equivalent width of most metal lines. This has  a (qualitatively) similar effect to increasing the flux ratio or decreasing the donor's effective temperature. The best-fit microturbulence is $v_{\rm turb}=2.5\,\rm km\,s^{-1}$, larger than typical value of $\sim 1\, \rm km\,s^{-1}$  for main-sequence stars. 

The spectra have per-pixel SNR between 500 and 2000, so full spectral fitting produces unreasonably small formal uncertainties (e.g., a few Kelvin in $T_{\rm eff}$). These dramatically underestimate the true uncertainties -- to the point of not being particularly useful -- because they do not account for systematics in the model spectra (e.g., imperfect linelists, assumption of LTE, 1D treatment of convection), systematics in the spectral fitting (e.g., assuming most metal abundances track Fe, imperfect normalization), or systematics in the data (e.g. imperfect masking of tellurics and DIBs, bad pixels, unrecognized contamination from emission lines, etc). This is a perennial issue for full spectral fitting in the high-SNR regime \citep[e.g.][]{Ness2015}. To obtain a more conservative estimate of the uncertainties, we varied the microturbulent velocity by $\pm 1\,\rm km\,s^{-1}$ around its fiducial value of $2.5\,\rm km\,s^{-1}$ (i.e., fixing it to 1.5 and 3.5 $\rm km\,s^{-1}$) and re-fit the other parameters. The $\pm$ uncertainties we report are half of the minimum-to-maximum range across the three fixed values of $v_{\rm turb}$, with the reported best-fit values corresponding to $v_{\rm turb}=2.5\,\rm km\,s^{-1}$. This approach leads to serviceable uncertainties that are more believable than the formal fitting errors, while still capturing the most important sources of uncertainty and the covariances between parameters.


\begin{figure}
    \centering
    \includegraphics[width=\columnwidth]{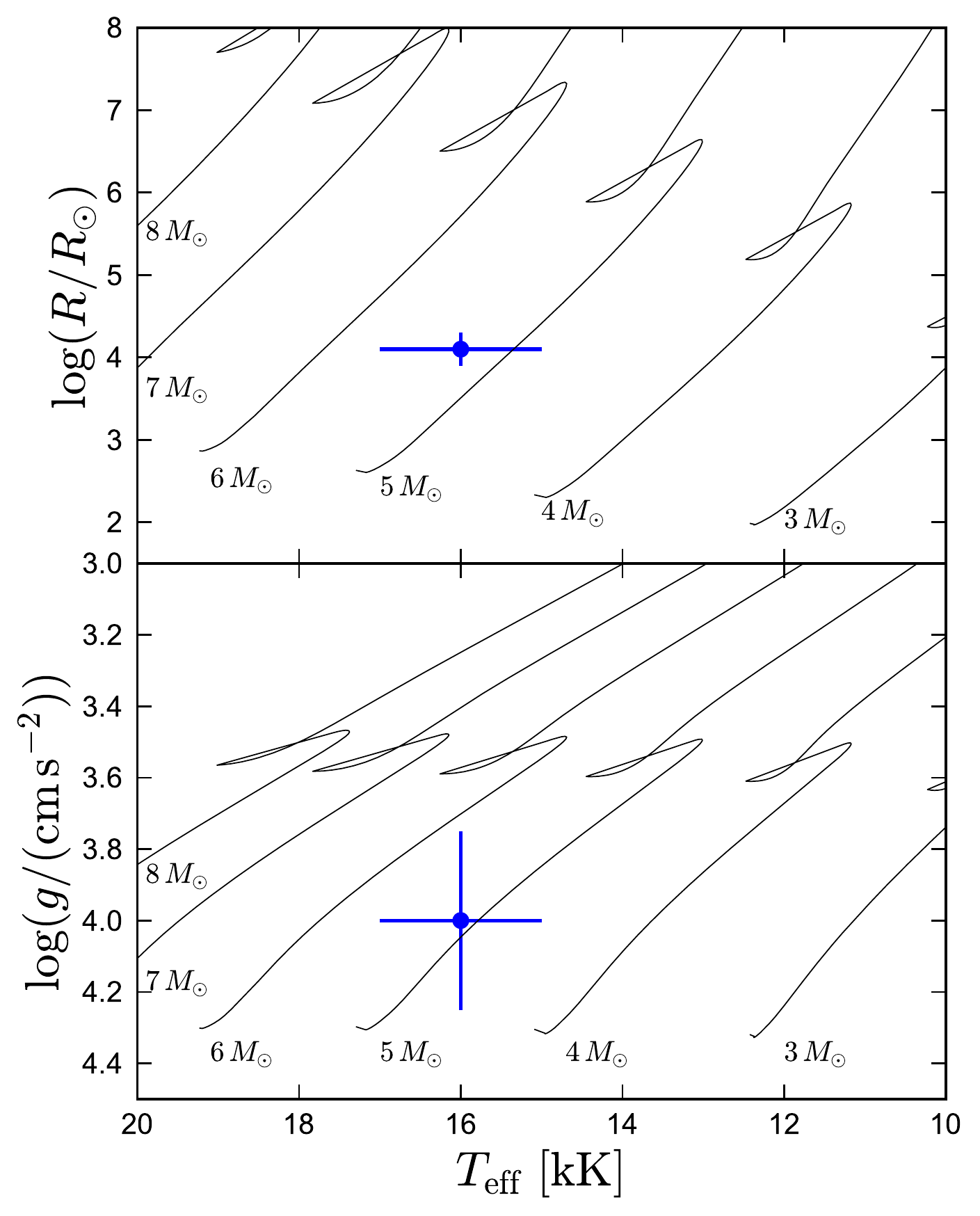}
    \caption{Measured parameters of the Be star in HD 15124 compared to MIST evolutionary tracks for single stars. Effective temperature and surface gravity are constrained spectroscopically; radius is constrained by the spectral energy distribution. The observed parameters imply a mass $M_{\rm Be}=5.3\pm  0.6 M_{\odot}$. }
    \label{fig:be_star_teff_logg}
\end{figure}

\subsection{Be star temperature, radius, and mass}
\label{sec:bemass}
Figure~\ref{fig:fluxed} shows the flux-calibrated spectrum and the predicted spectra of the best-fit model. The Be star's effective temperature and radius are only weakly sensitive to the properties of the donor, which contributes $\lesssim 20\%$ of the blue optical light. 

We compare the inferred parameters of the Be star to MIST evolutionary models \citep{Choi2016} for solar-metallicity main-sequence stars in Figure~\ref{fig:be_star_teff_logg}. To estimate the mass of the Be star, we construct a grid of evolutionary tracks with a mass spacing of 0.05 $M_{\odot}$, drawing 500 samples spaced uniformly in time from each track. We then retain samples with probability proportional to the likelihood of their effective temperature and radius given the measured constraints on the Be star's atmospheric parameters, approximating the likelihood function as a 2D Gaussian. This effectively selects points from the evolutionary tracks that are close to the Be star in temperature and radius. We use the resulting samples to estimate the mass and luminosity of the Be star, and to propagate forward the uncertainty in other parameters that depend on the Be star's properties. This yields a Be star mass $M_{\rm Be}=5.3\pm 0.6 M_{\odot}$.  Implicit in this modeling is the assumption that the Be star falls on a mass/radius/temperature relation similar to normal main-sequence stars (see Section~\ref{sec:inc}).

We note that we could also calculate $M_{\rm Be}$ directly from the measured surface gravity and radius. This leads to $M_{\rm Be} = 7.7\pm 4.3\,M_{\odot}$. This value is consistent with our inference from the evolutionary models, but its uncertainty is large due to the uncertainty in $\log g$.

\subsection{Donor temperature, radius, and mass}
\label{sec:donormass}

\begin{figure}
    \centering
    \includegraphics[width=\columnwidth]{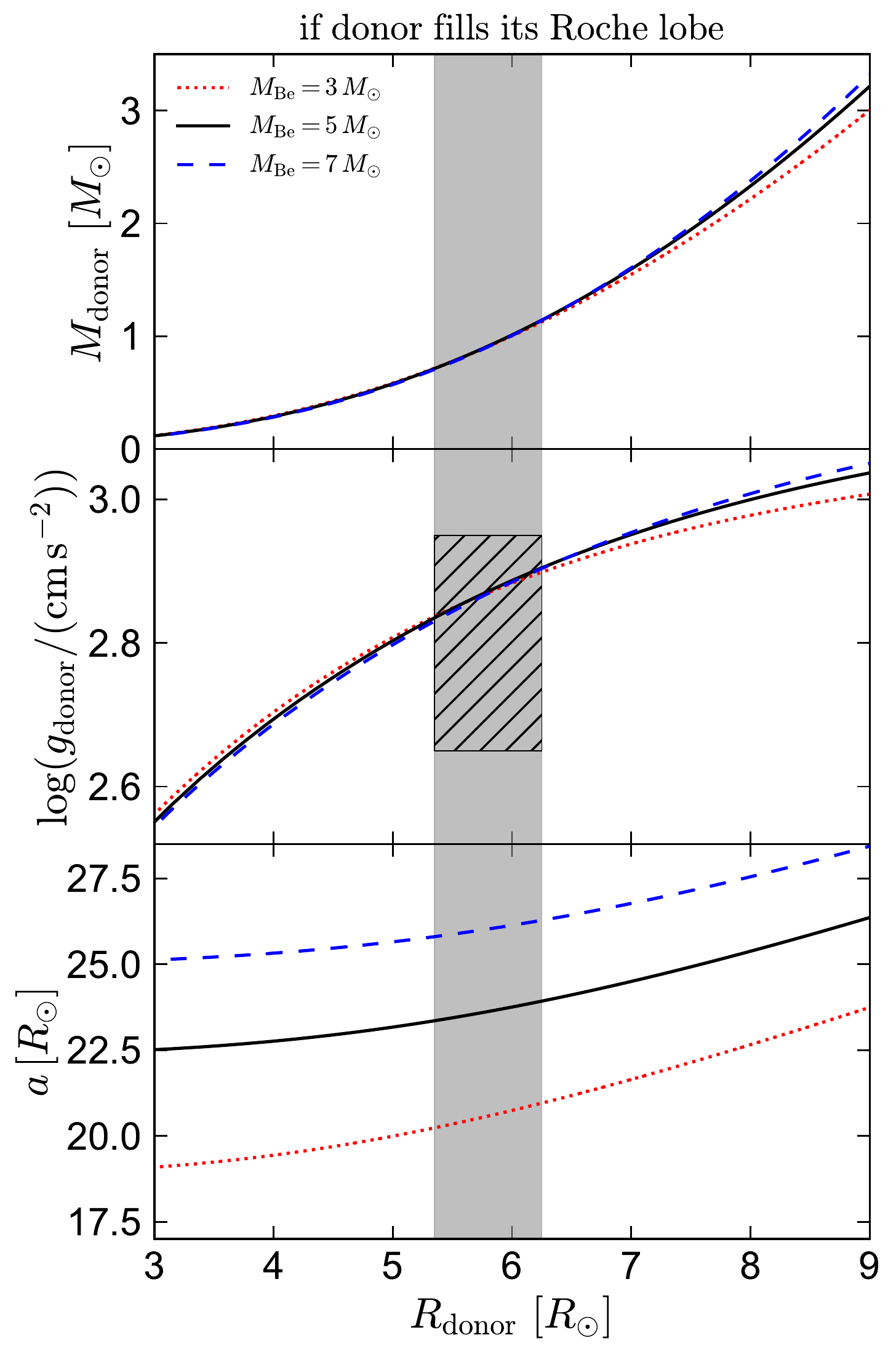}
    \caption{Donor mass, donor surface gravity, and binary semi-major axis predicted for different donor radii in HD 15124 if the donor fills its Roche lobe, as we argue it must. Shaded region shows the range of plausible donor radii, given our fits to the SED and spectrum. Hatched region shows the observational $\log g$ constraint. The donor mass and surface gravity are nearly independent of the mass of the Be star for any fixed donor radius; these quantities depend only on the (known) orbital period and the donor mass and radius. If the donor did {\it not} fill its Roche lobe, its mass, surface gravity, and semi-major axis would all be {\it larger} than these predictions. If it does, our constraint on $R_{\rm donor}$ implies $M_{\rm donor}=0.92\pm 0.22\,M_{\odot}$, $\log g_{\rm donor}=2.87\pm 0.04$, and $a=24\pm 0.8\,R_{\odot}$. }
    \label{fig:roche}
\end{figure}

The donor effective temperature and radius are also well-constrained by the spectra. The radius constraint comes from the normalization of the flux-calibrated spectrum and constraint on the optical flux ratio from the high-resolution spectrum. We find $R_{\rm donor} = 5.82\pm 0.45 R_{\odot}$ and $T_{\rm eff,\,\rm donor}=6900\pm 250\,\rm K$, with the flux ratio the dominant source of uncertainty. The spectroscopic surface gravity is $\log (g_{\rm donor}/{\rm cm\,s^{-2}})=2.8\pm 0.15$.

The mass of the donor can be constrained from the fact that it must fit within an orbit with $P_{\rm orb}=5.47$ days; i.e., the donor must be massive enough that its observed radius is not larger than its Roche lobe radius. Taking the expression for the Roche lobe radius from \citet{Eggleton_1983} leads to a lower limit on $M_{\rm donor}$ for any given $R_{\rm donor}$ that depends only very weakly on $M_{\rm Be}$. This constraint is shown in Figure~\ref{fig:roche}, and is essentially a reflection of the fact that all Roche lobe-filling stars at a given $P_{\rm orb}$ have nearly the same density, irrespective of their companion mass. 

With a lower limit on the donor mass in place from the Roche lobe constraint, an upper limit can be set in multiple ways. First, the spectroscopic $\log g=2.8\pm 0.15$ and radius constraints imply $M_{\rm donor} = (0.77 \pm 0.32) M_{\odot}$, or $M_{\rm donor}<1.09 M_{\odot}$, not much larger than the Roche lobe lower limit. Second, the amplitude of ellipsoidal variation depends strongly on the Roche lobe filling factor \citep[e.g.][their Figure 10]{El-Badry2021_pre_elms}. Based on the secondary minima present in the TESS light curve (Figure~\ref{fig:light_curves}), we find $R_{\rm donor}/R_{\rm Roche\,lobe}\gtrsim 0.93$. This translates to a donor mass only $\sim 25\%$ larger than when $R_{\rm donor}/R_{\rm Roche\,lobe} = 1$. For $R_{\rm donor} = 5.82$, requiring that $0.93 \leq R_{\rm donor}/R_{\rm Roche\,lobe} \leq 1$ constrains the donor mass to $0.92 \leq M_{\rm donor}/M_{\odot} \leq 1.15$. 

Since the light curve and spectroscopic $\log g$ suggest that the donor is close to Roche lobe-filling, we suspect that  mass transfer is still ongoing and the donor is semi-detached. This naturally explains the fact that the Be star has a circumstellar disk while rotating at a lower fraction of its critical rotation velocity than most Be stars (Section~\ref{sec:vsini}): it is still being spun-up by accretion. Moreover, all the evolutionary models we construct to match the system (Section~\ref{sec:mesa}) are still mass-transferring when they have temperature and surface gravity similar to the donor in HD 15124. We thus proceed under the assumption that the donor exactly fills its Roche lobe. This leads to a donor mass constraint of $M_{\rm donor}=0.92\pm 0.22\,M_{\odot}$. The uncertainty is dominated by the uncertainty on $R_{\rm donor}$, and thus would be only a factor of $\sim$1.5 larger if we did not treat the system as semi-detached.

\subsection{Inclination}
\label{sec:inc}
From the measured RVs of the donor, we calculate a mass function $f(M_{\rm Be})=0.23 \pm 0.01\,M_{\odot}$. This quantity represents the absolute minimum mass of whatever object the donor is orbiting. In particular, it is the mass that that object would have if the donor were a massless test particle and the system were viewed edge-on. For a finite donor mass or lower inclination, the implied mass is higher. 

We proceed under the assumption that the donor is orbiting the Be star, and interpret $f(M_{\rm Be})$ as a constraint on the Be star's mass. This assumption is not guaranteed to hold, since we have not measured RVs for the Be star to demonstrate the two stars are orbiting each other. It is in principle possible that the donor is orbiting another faint object (say, a white dwarf) and the Be star is a distant tertiary or a chance projection. However, we consider both scenarios unlikely. The chance projection scenario can be rejected because the probability of a close alignment of two bright stars with less than one arcsec separation is very small: following the approach used by \citet{El-Badry2021_wb}, we find a chance alignment probability of $\lesssim 10^{-6}$. The triple scenario with a faint close companion is less unlikely, but would not explain the observed reflection effect in the light curve (Figure~\ref{fig:light_curves}). We thus assume the donor is orbiting the Be star. We discuss the triple scenario with an outer Be star further in Appendix~\ref{sec:disc_triple}.

The observed mass function constrains the binary to a 2-dimensional surface in the space of donor mass, companion mass, and orbital inclination. This is illustrated in the bottom panels of Figure~\ref{fig:rvs}, which show the implied Be star mass for various combinations of donor mass and inclination. The relatively low mass function implies either a low companion mass or a low inclination (close to face-on). Given our constraints on the masses of both stars, the mass function implies an inclination $i = 23.2\pm 1.5$ degrees. We also consider the possibility that our spectroscopic constraint on the companion mass could be catastrophically wrong (e.g., if the Be star were a distant tertiary to an inner binary). In this case, the mass function would imply a companion mass of $0.92\pm 0.02 M_{\odot}$ for an inclination of 90 degrees, or  $1.15\pm 0.02 M_{\odot}$ for an inclination of 60 degrees. However, we consider such a scenario unlikely (Appendix~\ref{sec:disc_triple}).

Only about 8\% of randomly oriented orbits have inclinations as low as 23 degrees. It is thus worth considering whether there could be any serious systematic error affecting the inclination measurement. The inferred inclination depends mainly on the mass function and  the  mass of the Be star. The mass function is quite secure. The Be star's mass was inferred by comparing its spectroscopic temperature and radius to evolutionary models (Figure~\ref{fig:be_star_teff_logg}). This inference treats the Be star as a main sequence star and is thus somewhat less secure, because the Be star could in principle be out of equilibrium due to recent accretion. However, our MESA models (Section~\ref{sec:mesa}) suggest that departures from equilibrium for the accretor are modest, such that the accretor temperature and radius remain similar to those of a main-sequence star of the same mass. Even if our estimate of the Be star's mass were too large by a factor of 2 (which seems very unlikely), the implied inclination would remain low, $i \approx 33$ deg. We conclude that the inclination is genuinely low.

\subsection{Rotation velocities}
\label{sec:vsini}
We find projected rotation velocities $v\sin i= 20\pm 3\,\rm km\,s^{-1}$ for the donor and  $v\sin i= 95\pm 10\,\rm km\,s^{-1}$ for the Be star. The constraint for the donor comes primarily from metal lines in the PEPSI spectrum, while that for the Be star is from helium lines. If we assume the spin axes of both stars are aligned with the binary's orbital angular momentum vector -- a reasonable assumption for a close mass-transferring system -- then the constraint on the orbital inclination allows us to translate these into a synchronization parameter for the donor, $v_{{\rm rot}}/\left(2\pi R/P_{{\rm orb}}\right)=0.94\pm 0.17$; i.e., the donor is probably tidally synchronized. This validates our inferred inclination: a near Roche-filling star is expected to be tidally synchronized, and the measured $v\sin i$ would not correspond to synchronous rotation if it were not for the low inclination.

The implied critical rotation fraction for the Be star is $v_{\rm rot}/v_{\rm crit} = 0.60 \pm 0.08$. This is a factor of 2-3 larger than typical values for normal (non-emission line) B stars \citep{Abt2002, Huang2010}, but is also lower than the typical values of $\approx 0.8-0.9$ for classical Be stars \citep[][]{Townsend2004}. This suggests that the star has been spun up by accretion, but has not yet reached critical rotation. 

\subsection{Be star disk}
\label{sec:disk}
The velocity separation between the two peaks of the emission lines ranges from 250 to 350\,$\rm km\,s^{-1}$, while the separation between the line wings ranges 500 to 600 $\rm km\,s^{-1}$ (Figure~\ref{fig:pepsi}). If we assume that the emission is optically thin, then the separation between the peaks traces the Keplerian velocity at the outer edge of the disk, since for typical emissivity profiles, that is where most of the emission originates \citep[e.g.][]{Smak1969}. Given the inferred inclination, the peak-to-peak velocity separation would correspond to orbital disk velocities of $v_{\rm outer}\approx 380\pm 70\,\rm km\,s^{-1}$, corresponding to an outer disk radius of $R_{{\rm outer}}=GM_{{\rm Be}}/v_{{\rm outer}}^{2}\approx\left(7.0\pm3.5\right)R_{\odot}$, or in terms of the orbital separation, $R_{{\rm outer}}=\left(0.29\pm0.14\right)a$. This is consistent with the maximum stable disk radius of $R_{\rm max}\approx 0.42a$ for a binary with mass ratio $\approx 0.2$ \citep{Paczynski1977}, so the emission may trace material out to the largest stable streamline. We caution, however, that the emission is not necessarily completely optically thin, and the above does not hold if the line profile is significantly broadened by non-coherent scattering (i.e., absorption and re-emission at a different wavelength). Indeed, scattering must contribute to broadening at least the line wings, since their velocity separation would correspond to $(2.1\pm 0.6 R_{\odot})$ -- inside the Be star -- if it were purely Keplerian.

\subsection{Chemical abundances}
\label{sec:abundances}

\begin{table}
\caption{Photospheric abundances. We assume both stars have the same surface abundances. In practice, the constraint on He comes primarily from the Be star, while the constraints on all other elements are from the donor. Other metals track Fe, assuming the solar abundance pattern from \citet{Asplund2009}. Uncertainties in the 4th column also apply to the 1st column. }
\label{tab:abundances}
\begin{tabular}{lccc}
Element & $\log(n/n_{\rm tot})$ & $\log(n/n_{\rm tot})_{\odot}$ & $\log\left(\frac{n/n_{{\rm tot}}}{\left(n/n_{{\rm tot}}\right)_{\odot}}\right)$ \\ 
\hline
He      &   -0.51  &  -1.11  &   $0.6\pm 0.1$       \\
C       &   -5.91  &  -3.61  &   $-2.3\pm 0.2$      \\
N       &   -3.41 &  -4.21   &   $0.8 \pm 0.2$      \\
O       &   -4.15 &  -3.35   &   $-0.8 \pm 0.2$     \\
Fe      &   -4.49  &   -4.54 &   $0.05 \pm 0.1$     \\
\hline
\end{tabular}
\end{table}

\begin{figure*}
    \centering
    \includegraphics[width=\textwidth]{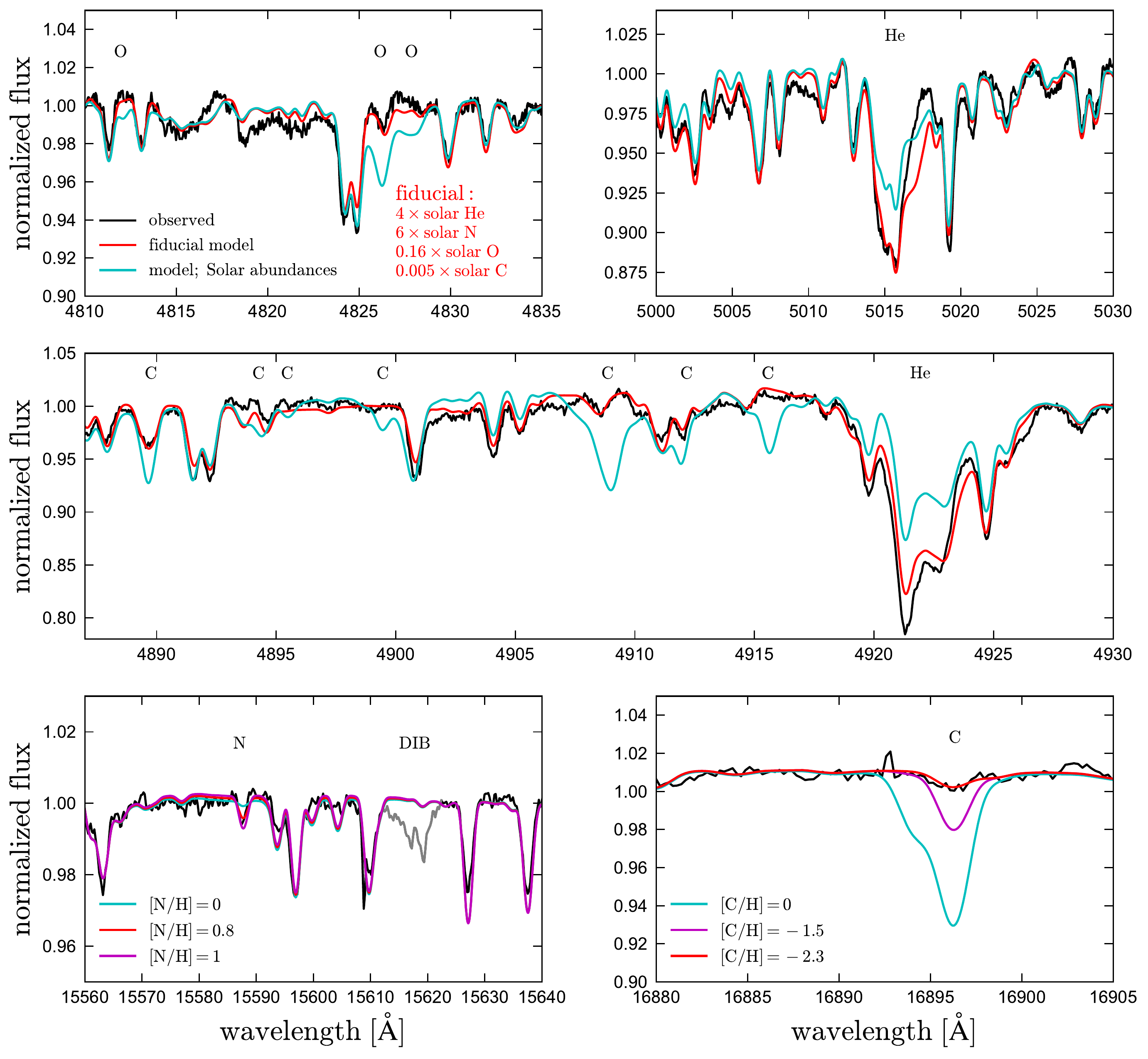}
    \caption{Spectral cutouts  highlighting evidence of CNO processing. Top two rows show PEPSI spectra; bottom row shows APOGEE spectra. In all panels, black line shows the observed spectra, and colored lines show composite model spectra with the parameters listed in Table~\ref{tab:system}. Helium lines are primarily from the Be star; all narrow lines are primarily from the donor, with continuum dilution from the Be star. Red line shows the best-fit model, the abundances of which are listed in the upper left panel. Cyan shows solar abundances, and magenta (bottom panels) shows other choices of N and C abundances. The spectra reveal strong enhancement of helium and nitrogen at the expense of carbon and oxygen, a sign of CNO processing. This suggests that the material currently on the surface of both stars was previously inside the convective core of the donor. }
    \label{fig:abundances}
\end{figure*}

The surface abundances (Table~\ref{tab:abundances}) reveal significant enhancement of helium and nitrogen, and depletion of carbon and oxygen. Figure~\ref{fig:abundances} compares cutouts of the high-resolution spectra to the best-fit model (cyan) and a model with the same atmospheric parameters and flux ratios but Solar abundances. The helium abundance is constrained by several strong lines in the PEPSI spectrum, and the carbon abundance by many lines in both the PEPSI and APOGEE spectra. The oxygen and nitrogen abundances are determined by a smaller number of weaker lines but still appear well-constrained. 

The best-fit model has a helium abundance by number that is a factor of 4 (0.6 dex) larger than the solar value. This corresponds to a surface helium mass fraction  $Y_{\rm He} = 0.63 \pm 0.08$. Carbon is depleted by 2.3 dex (more than a factor of 100), while oxygen is depleted by 0.8 dex and nitrogen is enriched by 0.8 dex. The carbon-to-nitrogen ratio is thus a factor of $\sim 1000$ larger than the solar value. The abundances of other elements appear normal. 

These abundances manifestly imply that the material currently on the surface of both stars was previously processed in the CNO cycle, which converts hydrogen to helium using C, N, and O as catalysis. In stars with masses of a few to a few tens of solar masses, CNO burning leads to a chemical equilibrium of C, N, and O in the convective core while the star is on the main sequence \citep[e.g.][]{Maeder2014}. Once the outer envelope of the donor is stripped off, this material is exposed in the photosphere. HD 15124 thus provides a rare opportunity to effectively look inside the core of an intermediate mass star. As we show in Section~\ref{sec:mesa}, the current surface abundances are in good agreement with theoretically predicted values in such a scenario.



\subsection{Galactic orbit}
\label{sec:orbit_mw}

\begin{figure*}
    \centering
    \includegraphics[width=\textwidth]{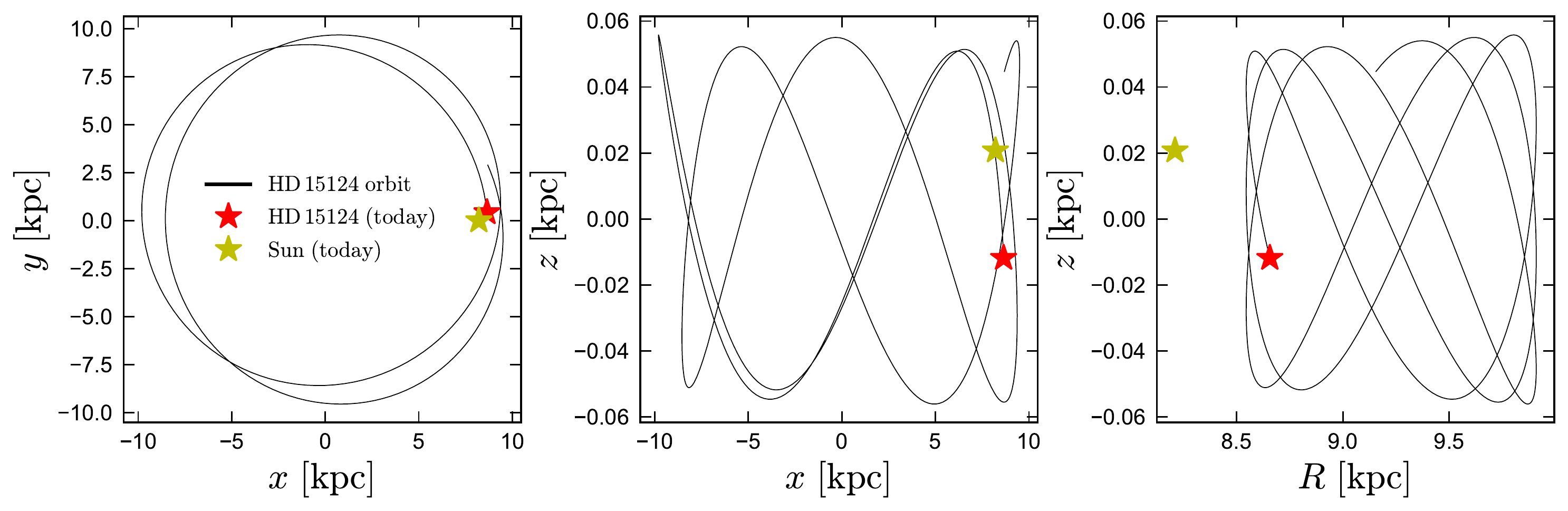}
    \caption{Galactic orbit of HD 15124, computed backwards for 500\,Myr from the measured astrometry and center-of-mass RV using the potential from \citet{McMillan2017}. Red and yellow stars show the present-day location of HD 15124 and the Sun. The orbit is typical for a young star in the Galactic thin disk.}
    \label{fig:orbit_mw}
\end{figure*}

To investigate the past trajectory of HD 15124, we used its {\it Gaia} astrometry and inferred center-of-mass radial velocity as starting points to compute its orbit backward in time for 500 Myr (though we note that the evolutionary age of the system is only 100-200 Myr) using \texttt{galpy} \citep{Bovy2015}, adopting the Milky Way potential from \citet{McMillan2017}. The result is shown in Figure~\ref{fig:orbit_mw}. The orbit appears typical of a thin-disk star, with excursions above the disk midplane limited to $\pm 50$\,pc. We verified that this conclusion is robust to changes in the assumed potential and to varying the astrometry and RV within the observational uncertainties.

\section{Evolutionary history of the HD 15124 system}
\label{sec:mesa}
\subsection{BPASS library search}

To understand the possible formation pathways and future evolution of HD 15124, we began by searching the BPASS (v2.2; \citealt{Eldridge2017}; \citealt{Stanway2018}) library of binary evolution calculations for models that go through a phase similar to the binary's current state.
We searched the $Z=0.02$ grid for models that at any point in their evolution satisfy the following constraints: 
\begin{itemize}
    \item $5900 < T_{\rm eff,\,donor}/{\rm K} < 7900$
    \item $4 < R_{\rm donor}/R_{\odot} < 8$
    \item $4 < M_2/M_{\odot} < 10 $
    \item $0 < P_{\rm orb}/{\rm days} < 10$
\end{itemize}
Here $M_2$ is the mass of the initially less-massive star. This search yielded 18 models, with initial primary masses ranging from 3.5 to 5 $M_{\odot}$, initial mass ratios ranging from 0.5 to 0.9, and initial periods of 1.6 to 2.5 days. Their evolutionary history is qualitatively quite similar to the models explored to explain the HR 6819, LB-1, and NGC 1805 BH1 systems \citep[e.g.][]{Eldridge2020, Bodensteiner2020, El-Badry2021, El-Badry_burdge2021, Stevance2021}. In all the selected models, mass transfer begins while the donor is still on the main sequence (``case A'').   While these models do come close to the observed properties of HD 15124, they all predict donor masses and radii on the high side of the observed values (median $M_{\rm donor}\approx 1.7\,M_{\odot}$ and $R_{\rm donor}\approx 7.1\,R_{\odot}$). The surface abundances of the donors in these models suggest that they are less stripped than HD 15124 (median $Y_{\rm He}=0.31$, only slightly enriched compared to the initial value of 0.28). 

Models with lower donor masses and higher $Y_{\rm He}$ do exist in the BPASS library, but these have longer periods. This is a consequence of the way mass and angular momentum transfer are implemented in the BPASS models: mass transfer is assumed to be fully conservative, unless it occurs on a timescale shorter than the accretor's thermal timescale (e.g., $\dot{M} > M_{\rm accretor}/t_{\rm KH,\,accretor}$, where $t_{\rm KH}$ is the Kelvin-Helmholtz time). This leads to rapid widening of the orbit once the donor is the lower-mass star (e.g. Equation~\ref{eq:p_final} below). To investigate the evolution of the system in more detail and explore models with different treatments of angular momentum loss, we calculated additional models.

\subsection{MESA calculations}

We calculated a small grid of binary evolution models using MESA \citep[Modules for Experiments in Stellar Astrophysics, version \texttt{r15140};][]{Paxton_2011, Paxton_2013, Paxton_2015, Paxton_2018, Paxton_2019}. We primarily explored initial conditions similar to those identified by our BPASS search, but also experimented with varying the initial conditions and input physics. Unlike the models in the BPASS library, which only follow the primary star with detailed calculations, MESA simultaneously solves the 1D stellar structure equations for both stars, while accounting for mass and angular momentum transfer using simplified prescriptions. 

We assumed an initial composition of $X=0.7, Y=0.28, Z=0.02$ for both stars. Opacities are taken from OPAL  \citep{Iglesias1996} at $\log(T/{\rm K}) \geq 3.8$ and from \citet{Ferguson2005} at $\log(T/{\rm K}) < 3.8$. Nuclear reaction rates are from \citet{Angulo1999} and \citet{Caughlan1988}. We use the \texttt{pp\_and\_cno\_extras} nuclear network, which includes $^{1,2}\rm H$, $^{3,4}\rm He$, $^{7}\rm Li$, $^{7}\rm Be$, $^{8}\rm B$, $^{14}\rm N$,  $^{14, 16}\rm O$, $^{19}\rm F$, $^{18,19,20}\rm Ne$, and $^{22,24}\rm Mg$. We use the \texttt{photosphere} atmosphere table, which uses atmosphere models from \citet{Hauschildt1999} and \citet{Castelli2003}. 

The \texttt{MESAbinary} module is described in \citet{Paxton_2015}. We used the \texttt{evolve\_both\_stars} inlists in the MESA test suite as a starting point for our calculations, and most inlist parameters are set to their default values. Both stars are evolved simultaneously, with the response of the secondary to accretion taken into account. Roche lobe radii are computed using the fit of \citet{Eggleton_1983}. Mass transfer rates in Roche lobe overflowing systems are determined following the prescription of \citet{Kolb_1990}. 
The orbital separation evolves such that the total angular momentum is conserved when mass is lost through winds or transferred to a companion, as described in \citet{Paxton_2015}. 

We experimented with both conservative and non-conservative mass transfer. As with the models from the BPASS library, we were unable to match the current properties of HD 15124 (orbital period, donor mass and radius, and significant surface helium enrichment) with models in which mass transfer is fully conservative (meaning that all mass lost by the primary is accreted by the secondary). The ``problem'' is the short orbital period. When mass transfer is conservative, there is a simple relation between the initial and current orbital periods and mass ratios \citep[e.g.][]{Soberman1997}:
\begin{align}
    \label{eq:p_final}
    P_{\rm orb}(t)=P_{\rm orb,\,init}\times \left(\frac{M_{\rm donor,\,init}}{M_{\rm donor}(t)}\times\frac{M_{\rm accretor,\,init}}{M_{\rm accretor}(t)}\right)^{3}.
\end{align}
Here $P_{\rm orb}(t)$, $M_{\rm donor}(t)$, and $M_{\rm accretor}(t)$ are the current orbital period and donor and accretor masses, while values with the subscript ``init'' are their values at the onset of mass transfer. Given the current $M_{\rm donor}\approx 0.92\,M_{\odot}$ and $M_{\rm accretor}\approx 5.3\,M_{\odot}$, the total initial mass would need to be about $6.2 M_{\odot}$ if mass transfer were fully conservative. To avoid the formation of a common envelope or contact binary, the initial masses are required to have been relatively similar; e.g., $M_{\rm accretor,\,init}/M_{\rm donor,\,init} \gtrsim 0.6$. Taking, as a representative example, $P_{\rm orb,\,init} = 2\,\rm days$ and $M_{\rm donor,\,init}=3.5 M_{\odot}$ and $M_{\rm accretor,\,init}=2.7 M_{\odot}$, Equation~\ref{eq:p_final} yields a current period $P_{\rm orb}\approx 15$ days, significantly longer than the observed value of 5.5 days. In order to match the observed period with these masses, Equation~\ref{eq:p_final} requires a shorter initial period. But in this case, the donor would overflow its Roche lobe early in its main-sequence evolution, leading to qualitatively different evolution (Section~\ref{sec:varying_parameters}).

When mass-transfer is {\it not} conservative, the orbital period evolution depends on the angular momentum of the material that escapes. \texttt{MESABinary} implements the ``$\alpha\beta\gamma\delta$'' formalism \citep[see][]{Huang1963, vandenHeuvel1994, Soberman1997, Tauris_2006} for non-conservative mass loss, in which a fraction $\alpha_{\rm ML}$ of the mass lost by the donor is ejected in a fast wind from the donor, a fraction $\beta_{\rm ML}$ is ejected in a fast wind from the accretor, and a fraction $\delta_{\rm ML}$ is lost from a circumbinary ring with radius $\gamma_{\rm ML}^2 \times a$. In our fiducial model, we fix $\alpha_{\rm ML}=\beta_{\rm ML}=0$, $\delta_{\rm ML} =0.2$, and $\gamma_{\rm ML}=\sqrt{2}$; that is, we assume 20\% of the mass leaving the donor is eventually lost from the binary through a circumbinary ring of radius $2a$, while the rest is transferred to the companion. Compared to the case of conservative mass loss, this tends to reduce the binary's orbital period, because the mass that is lost from the circumbinary ring has higher specific angular momentum than the binary itself. 

In reality, $\alpha_{\rm ML}$, $\beta_{\rm ML}$, $\gamma_{\rm ML}$, and $\delta_{\rm ML}$ are expected to be functions of the mass transfer rate and other instantaneous properties of the binary, and thus should vary with time. Ab-initio constraints on these parameters are scarce, and we do not attempt to infer values for them based on one system. For our purposes, the choice of $\delta_{\rm ML}=0.2$ and $\gamma_{\rm ML}=\sqrt{2}$ should be regarded as an effective model of a more complicated mass transfer process; likely, one of several possible combinations of $\alpha_{\rm ML}$, $\beta_{\rm ML}$, $\gamma_{\rm ML}$, and $\delta_{\rm ML}$ that can produce the right amount of angular momentum loss to match the observed period. We discuss how varying the parameters of the MESA model from our fiducial values changes the calculation in Section~\ref{sec:varying_parameters}.

\begin{figure*}
    \centering
    \includegraphics[width=\textwidth]{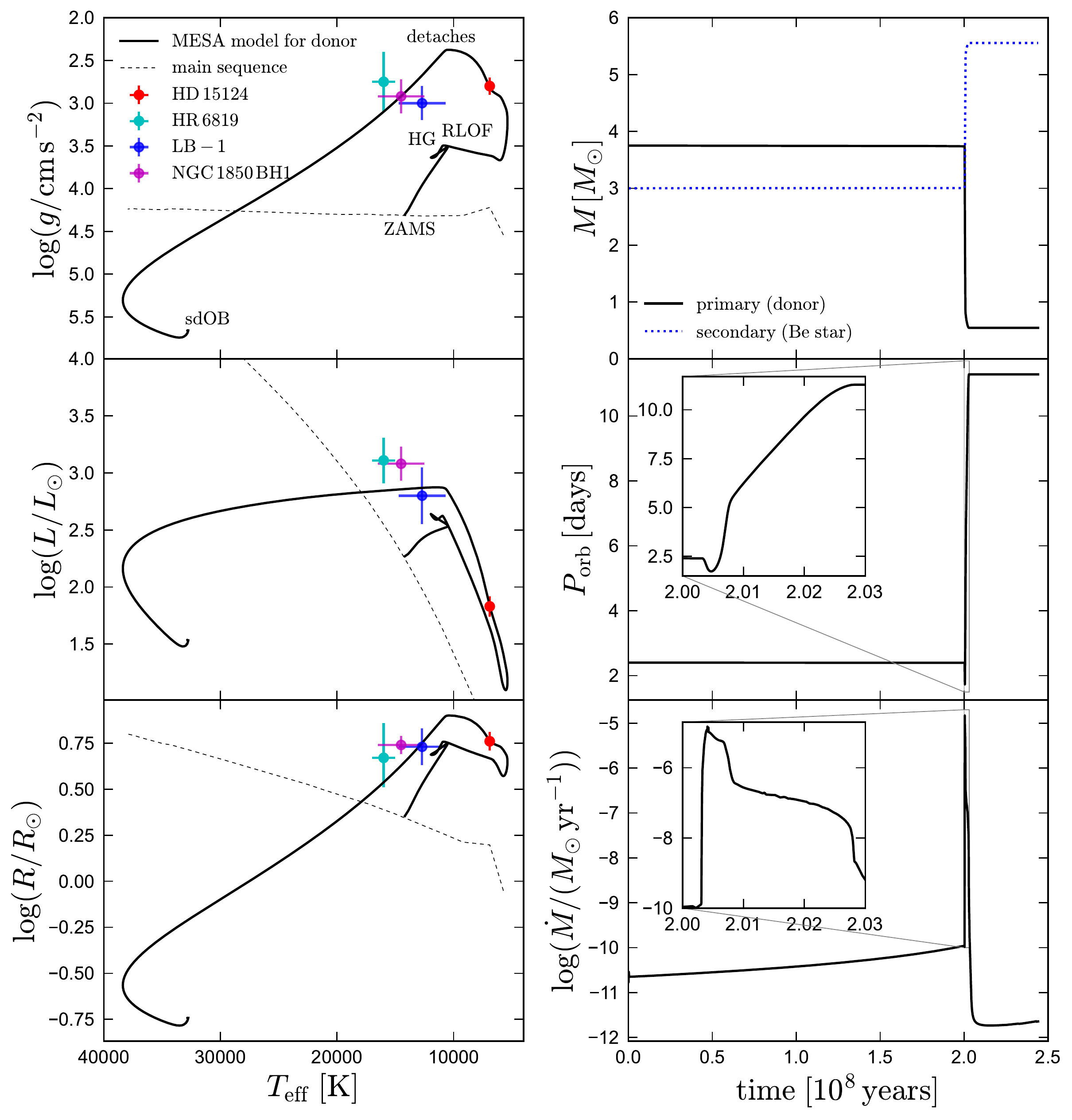}
    \caption{MESA binary evolution calculations for a system similar to HD 15124. Let panels show the evolution of the primary (i.e., the donor) in a binary with initial masses of 3.75 and 3 $M_{\odot}$ and initial period of 2.4 days. During its main-sequence lifetime, the primary evolves from the zero-age main sequence (ZAMS) to the Hertzsprung gap (HG).  Roche lobe overflow (RLOF) occurs as the star is crossing the HG and is followed by a brief ($\sim 3\times 10^5$ years) period of thermal-timescale mass transfer. The star cools at near-constant radius during this period, leading to a factor-of-40 drop in luminosity. The donor reaches $T_{\rm eff}\sim 7000\,\rm K$ and $\log g\sim 2.8$ (similar to HD 15124) at $P_{\rm orb}= 6-7$ days, when its mass loss rate is $\dot{M} \sim 10^{-7} \,\rm M_{\odot}\,yr^{-1}$. Mass transfer ceases after $\sim$2 Myr, when the donor has lost most of its envelope. The donor then contracts and heats up, settling as a core helium burning sdO/B star. During this contraction, the model's parameters are similar to those of the bloated stripped stars in HR 6919, LB-1, and NGC 1805 BH1, systems with similar evolutionary histories. The calculation ends after another $\sim$50 Myr, when the Be star evolves and overflows its Roche lobe.}
    \label{fig:mesa_evol}
\end{figure*}

Figure~\ref{fig:mesa_evol} shows the evolution of the initially more massive star in a MESA model that matches the present-day properties of HD 15124, which we adopt as our fiducial model. The initial component masses are 3.75 and 3.0 $M_{\odot}$, and the initial period is 2.4 days. Mass transfer begins as the primary is crossing the Hertzsprung gap and initially proceeds on its thermal timescale, leading to a maximum $\dot{M}$ of $\approx 10^{-5}\,M_{\odot}\,\rm yr^{-1}$. Here the donor rapidly cools, and its luminosity falls by more than a factor of 10. During this phase, the deeper layers of the donor are expanding to adjust to the decrease in total mass. The energy needed for this expansion comes at the expense of the surface luminosity, such that the nuclear luminosity significantly exceeds the surface luminosity.

Once the donor has lost most of its envelope, the mass transfer rate falls to $\dot{M} \simeq 10^{-7}\,M_{\odot}\,\rm yr^{-1}$ and the donor begins to expand. By now the donor has become the lower-mass star, so mass loss widens the orbit. This leads to a semi-detached system whose evolution is governed by expansion of the evolving donor. During this period, the model resembles HD 15124, with $M_{\rm donor} \approx 0.8\,M_{\odot}$, $T_{\rm eff, donor} \approx 7000\,\rm K$, and $\log(g_{\rm donor})\approx 2.8$.

\begin{figure}
    \centering
    \includegraphics[width=\columnwidth]{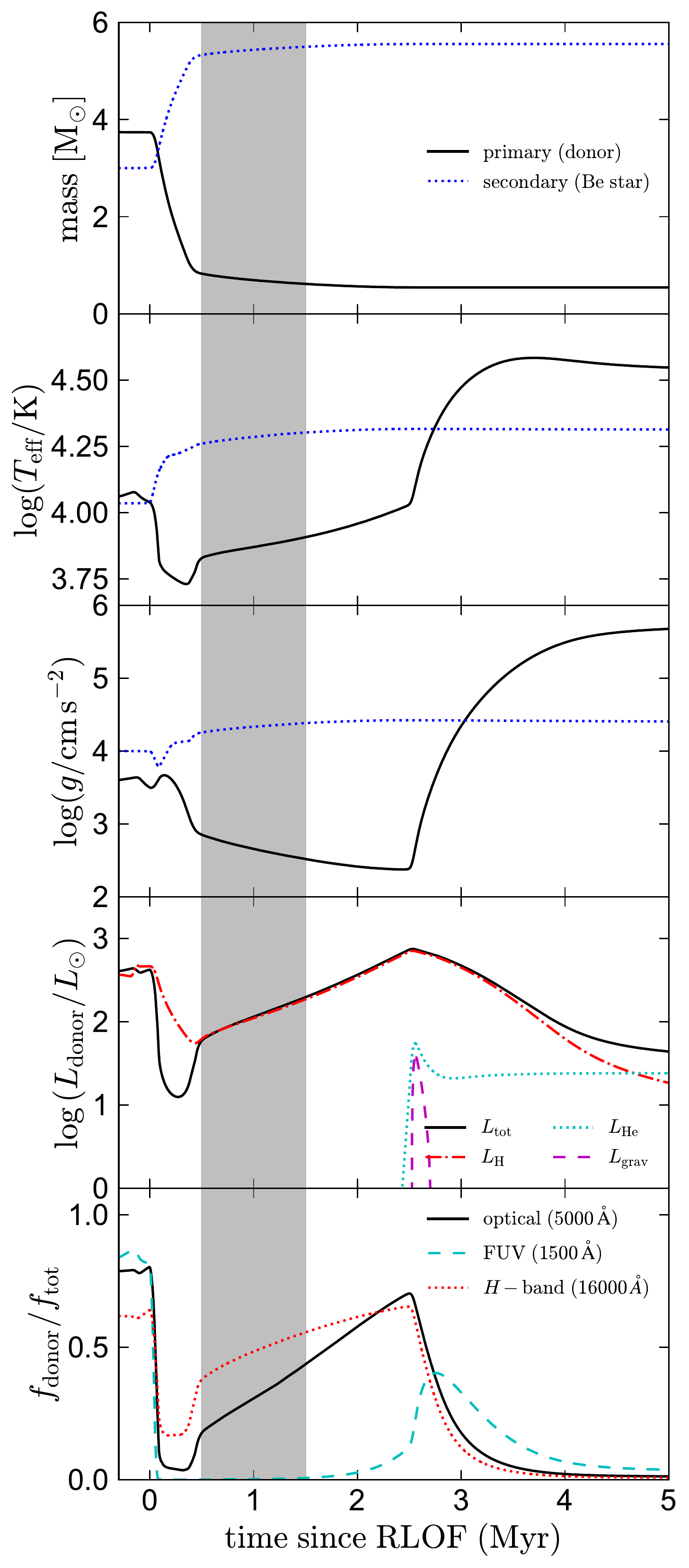}
    \caption{Time-evolution of the fiducial MESA model, zoomed-in on the period of interaction. Mass-transfer begins at $t=0$ and ends at $t=2.5\,\rm Myr$, after which the donor heats up and contracts. Most of the mass loss occurs in the first 0.5 Myr. Shaded region indicates when the donor temperature and surface gravity are similar to HD 15124, such that our APOGEE search would detect the companion. Third panel shows the donor luminosity (black) and luminosity of hydrogen burning (red), helium burning (cyan) and gravitational contraction (magenta). The luminosity is currently dominated by shell hydrogen burning but is predicted to become dominated by core helium burning within a few Myr. Bottom panel shows the fraction of the total flux contributed by the donor at optical, UV, and NIR wavelengths.}
    \label{fig:mesa_zoom}
\end{figure}

The evolution of the model during this period is shown in more detail in Figure~\ref{fig:mesa_zoom}. The period during which our APOGEE search would be sensitive to the donor lasts $\approx 1$ Myr, after which the donor becomes too hot to be detectable via its metal lines in the IR (Appendix~\ref{sec:sensitivity_appendix}). The binary remains semi-detached for another $\approx 1$\,Myr, until the donor's hydrogen envelope mass is reduced to $\sim 0.1\,M_{\odot}$. At this point, the donor begins to contract and heat up, eventually settling at a radius of $\approx 0.2\,R_{\odot}$ and effective temperature of $\approx 35,000\,\rm K$, typical values for a sdO/B star \citep[][]{Heber2016}.

The luminosity of the model during the observed phase and the first $\approx 2$\,Myr of contraction is powered by shell burning of the remaining hydrogen envelope. Core helium burning begins just as the donor begins to contract and dominates the luminosity after $\approx 2$\,Myr. The lifetime of the subsequent core helium burning phase would be of order 100 Myr, but in most of the models we explore, the companion finishes its main sequence evolution and overflows its Roche lobe first. This is likely to lead to a period of common envelope evolution, but we stop the calculation here.

In Figure~\ref{fig:mesa_evol}, we also compare the model to the donor stars in LB-1, HR 6819, and NGC 1850 BH1. These systems can all be described by a qualitatively similar evolutionary scenario in which an intermediate-mass primary overflows its Roche lobe near the end of its main-sequence evolution, goes through an Algol phase, and eventually contracts to become an sdO/B star. The details are, however, different in each system. LB-1 and HR-6819 have significantly longer orbital periods (79 and 40 days) and are thus better described by models with more conservative mass transfer. 
HD 15124 has a shorter orbital period than any known classical Be star binaries. Several Algol-type binaries with qualitatively similar evolutionary histories and similarly short periods are known \citep[e.g.][]{Tupa2013, Rosales2021}, but the donors in these systems likely have higher masses and are somewhat less evolved than HD 15124. The orbit of HD 15124 is expected to widen as mass transfer continues. The fiducial MESA model predicts a final period of 12 days, but the true value may be larger because mass transfer is expect to become more conservative as $\dot{M}$ decreases, an effect not included in the model.

\begin{figure}
    \centering
    \includegraphics[width=\columnwidth]{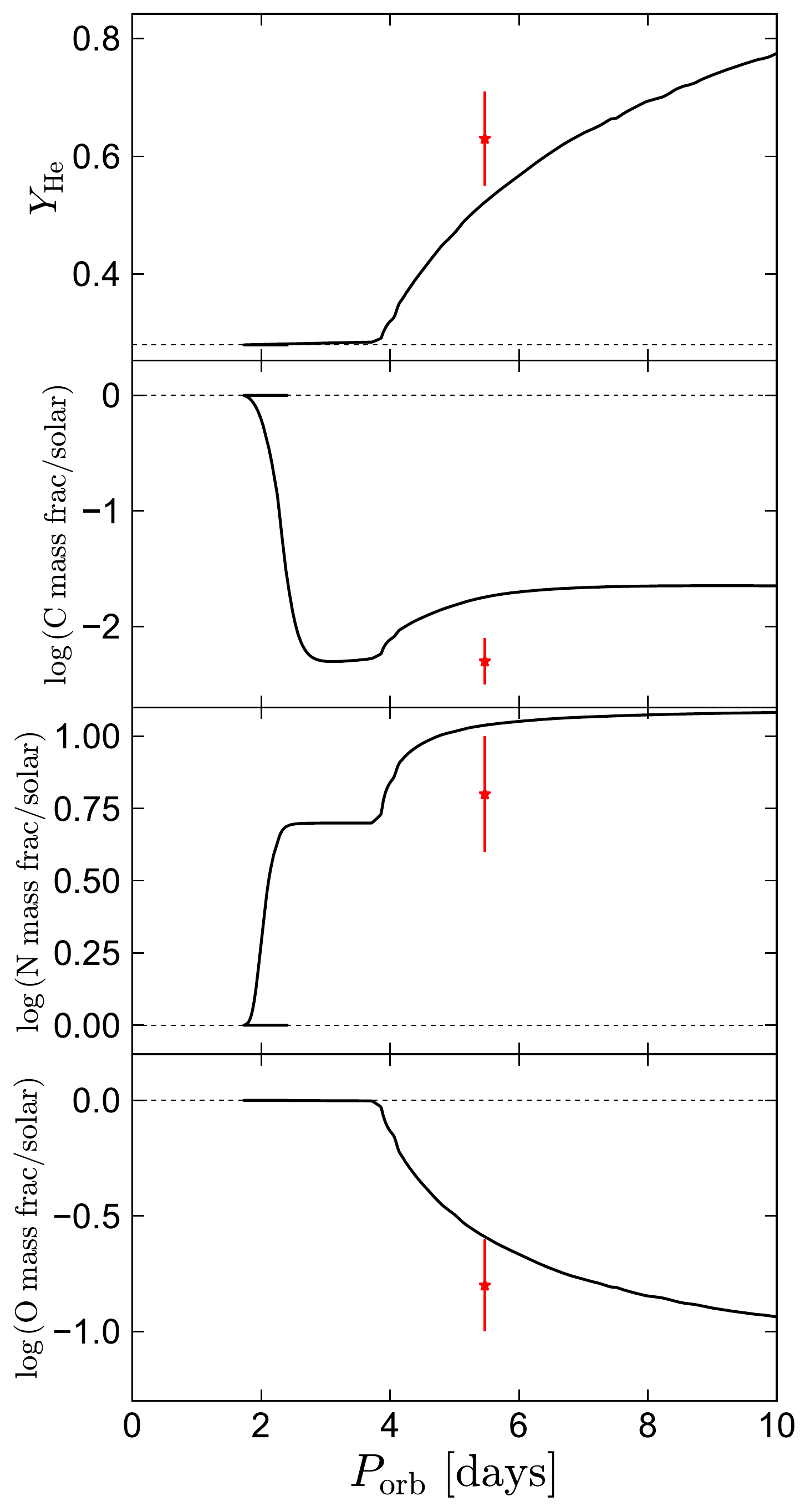}
    \caption{Surface abundances of HD 15124 and the MESA model from Figure~\ref{fig:mesa_evol}. All panels show mass fractions, and dashed horizontal lines show Solar values. The observed abundances of helium, carbon, nitrogen, and oxygen are all highly discrepant from the solar abundance pattern: helium and nitrogen are strong enriched, and carbon and oxygen are depleted. The model reproduces the observed abundances reasonably well. In the model, the current surface of the donor was previously inside the donor's convective core during CNO burning.}
    \label{fig:mesa_abun}
\end{figure}
\begin{figure*}
    \centering
    \includegraphics[width=\textwidth]{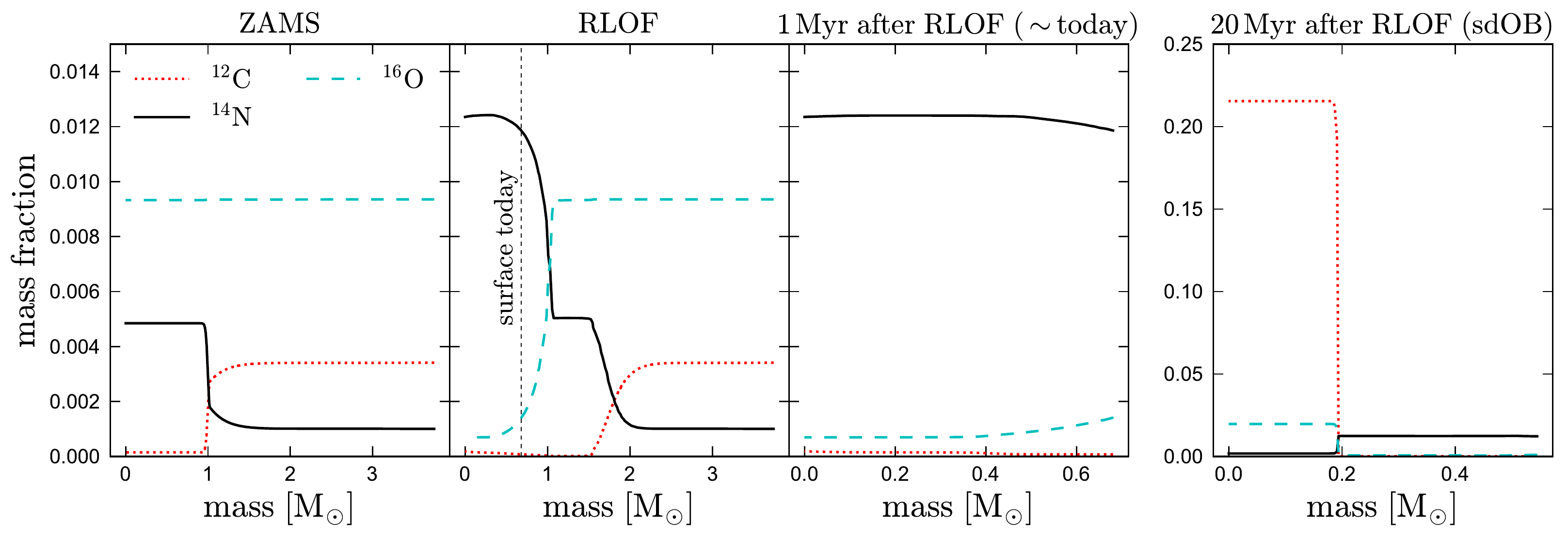}
    \caption{Profiles of CNO mass fractions in the donor, for the MESA model shown in Figure~\ref{fig:mesa_zoom}. The four panels show the zero-age main sequence, onset of Roche lobe overflow, 1 Myr later (corresponding roughly to the current state of HD 15124), and core helium burning. The horizontal axis changes between panels as mass is removed from the donor. Vertical line in panel 2 marks the approximate current surface of the donor. The surface of the star today fell within the convective core of the donor at RLOF, whose composition is heavily affected by CNO processing. }
    \label{fig:mesa_profiles}
\end{figure*}

Figure~\ref{fig:mesa_abun} shows the evolution of surface abundances of the donor in the MESA model as a function of orbital period. These are initially solar but reveal increasing levels of CNO processing as the period increases and deeper layers of the donor's interior are exposed. Figure~\ref{fig:mesa_profiles} shows the interior abundance profiles of the donor at several snapshots of its evolution. CNO ratios at the surface remain at the solar value until initial Roche lobe overflow, when the core is already heavily processed. Helium and oxygen are respectively enriched and depleted only in the central $\approx 1\,M_{\odot}$, while nitrogen and carbon are enriched and depleted out to $2\,M_{\odot}$. The observed surface abundances thus provide an additional constraint on how much of the donor's outer envelope has been stripped. 

\subsection{Varying model parameters}
\label{sec:varying_parameters}

\begin{figure*}
    \centering
    \includegraphics[width=\textwidth]{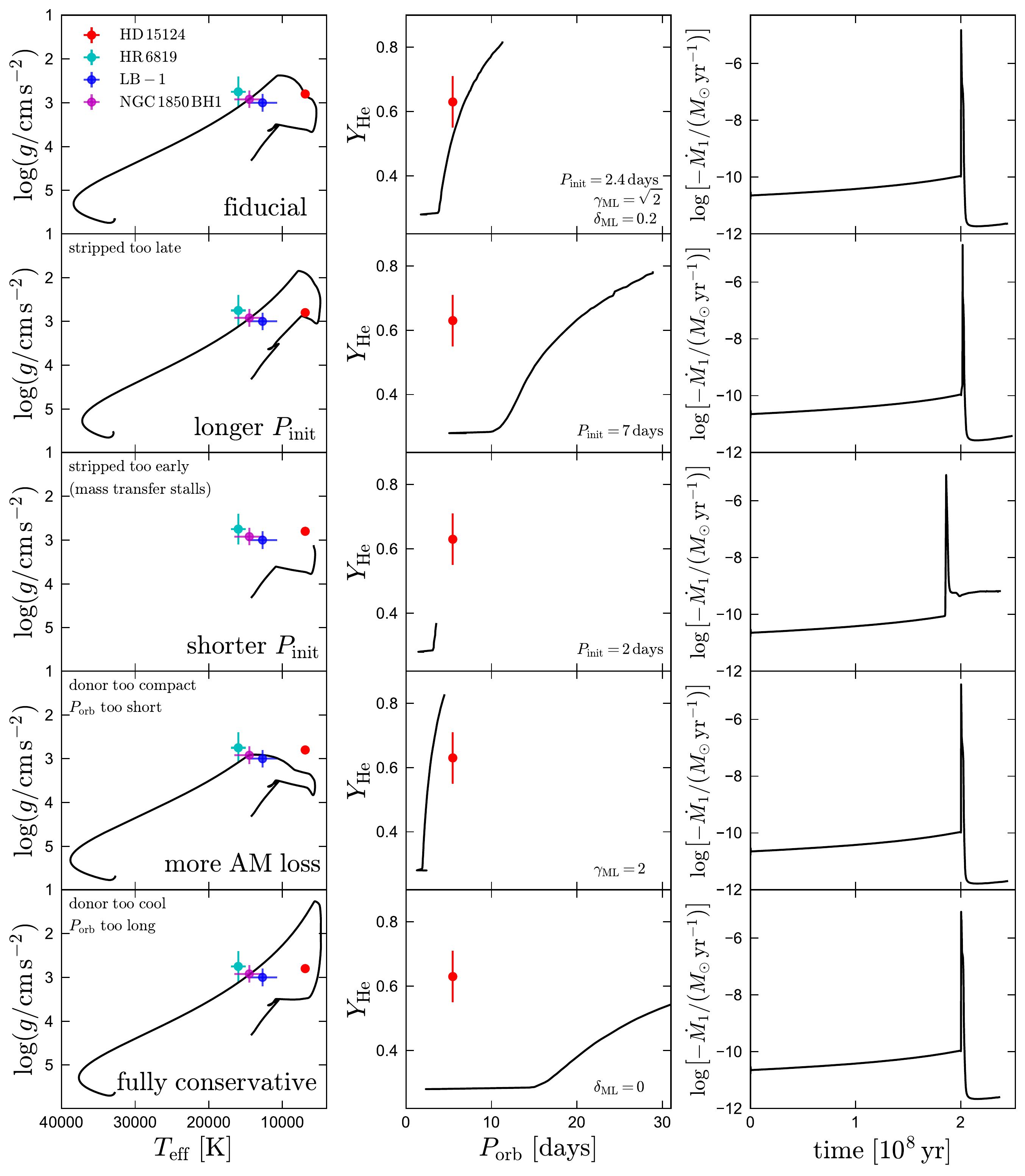}
    \caption{Effects of varying MESA model parameters. Top panel shows the fiducial model. Second panel shows a longer initial period, which causes mass transfer to begin later, such that significant helium enrichment is not predicted until longer periods than observed.  Third panel shows a shorter initial period. In this case, mass transfer begins before core hydrogen exhaustion in the donor but then stalls until the now-more-massive secondary evolves and overflows its Roche lobe. Third panel shows a larger value of $\gamma_{\rm ML}$, such that the specific angular momentum of matter lost from the binary is higher. This leads to more rapid stripping and a shorter orbital period. Bottom panel shows fully conservative mass transfer. In this case, the donor does not reach the observed degree of stripping as quantified by $Y_{\rm He}$ until a much longer orbital period. }
    \label{fig:mesa_vars}
\end{figure*}

Here we discuss how the behavior of the MESA models changes as we vary the initial conditions and model ingredients. Some representative models are shown in Figure~\ref{fig:mesa_vars}.

\begin{enumerate}
    \item {\it Larger or smaller initial period}: A moderately longer initial period ($3 \lesssim P_{\rm init}/{\rm day} \lesssim 20$, with all else fixed) causes mass transfer to begin farther up the subgiant branch, when the donor is larger. The model's evolution is qualitatively similar to in our fiducial model, but the current orbital period is longer than observed. At even longer initial periods ($20 \lesssim P_{\rm init}/{\rm day} \lesssim 150$), mass transfer begins when the donor is a giant, presumably leading to an episode of common envelope evolution (which cannot be properly followed in our calculations). This could potentially produce a short period as in HD 15124 but likely would not end with a donor that is still Roche lobe-filling. Even longer initial periods can produce stable mass transfer when the donor is an AGB star (stabilized by its larger core mass), but these also produce periods much longer than observed.
    
    A shorter initial period ($P_{\rm init} \lesssim 2$ days) causes mass transfer to begin while the donor is still on the main sequence. Mass transfer is initially similar to in the fiducial model. However, because the donor has not yet terminated its nuclear evolution, it shrinks back within its Roche lobe once it becomes the lower-mass star, and further evolution occurs on a (now longer) nuclear timescale. Eventually, the now-more-massive secondary evolves off the main sequence and overflows its Roche lobe. Most of the BPASS models we explored follow this type of evolution, but they do not track the response of the secondary to accretion and thus do not capture its more rapid subsequent evolution.
    
    \item {\it Higher or lower primary mass}: A higher initial primary mass is possible if (a) the secondary mass was also higher (because a too-unequal mass ratio leads to formation of a common envelope; see below) and (b) the mass transfer efficiency was lower, to avoid producing a too-massive Be star. Under these conditions, a higher initial primary mass generally leads to a higher current mass for the bloated core. A significantly lower initial primary mass is ruled out, since the total mass of the binary today must exceed $\sim 6\,M_{\odot}$.
    
    All the MESA models we found plausibly compatible with the current properties of HD 15124 produce helium cores with masses $0.5\lesssim M_{\rm He\,core}/M_{\odot}\lesssim 1$, such that the donors will become core helium burning sdO/B stars. Lower-mass analogs of the system produce donors that will contract directly to become helium white dwarfs \citep[e.g.][]{El-Badry2021_j0140, Miller2021}. Higher-mass systems produce more massive and hotter sdO/B stars \citep{Gotberg2017}, which eventually can explode and leave behind a Be X-ray binary \citep[e.g.][]{Townsend2020, Klement2021_interferometry}.
    
    \item {\it Larger or smaller mass ratio}: Mass transfer is less stable (i.e., more likely to result in a common envelope) when the ratio $M_{\rm donor}/M_{\rm accretor}$ is large. Models with highly unequal initial masses $(M_2/M_1 \lesssim 0.6)$ can in principle be evolved with MESA to produce systems like HD 15124 and similar binaries at more evolved stages \citep[e.g.][]{Bodensteiner2020}, but this entails evolving the model through a contact phase, which is not reliably captured by the models. Mass ratios closer to 1 can produce qualitatively similar evolution to our fiducial model. For initial mass ratios close to 1, the duration of the predicted Be + sdO/B phase becomes shorter, because the secondary leaves the main sequence sooner. 
    \item {\it Changes in mass transfer prescription}: If mass loss is conservative, the orbital period increases too rapidly to reproduce the observed value (Equation~\ref{eq:p_final}). This can produce wider systems like LB-1 and HR 6819. If mass loss is less conservative or removes more specific angular momentum than in our fiducial model, the orbital period will be shorter, and the donor will be fully stripped before it can ascend up the subgiant branch. This can produce a compact mass-transferring system like NGC 1850 BH1.

\end{enumerate}

\section{Implications for Be star formation channels}
\label{sec:discussion}

Once the donor in HD 15124 becomes an sdO/B star, it will be outshone by the Be star and very difficult to detect at any wavelength (bottom panel of Figure~\ref{fig:mesa_zoom}). Its mass at that time is predicted to be $\approx 10$ times lower than that of the Be star, so the expected RV shifts of the Be star will be only a few $\rm km\,s^{-1}$ -- too low to reliably detect in a Be star with high rotation velocity and disk-driven spectral variability. The lifetime of the Be + sdO/B phase is expected to be much longer than that of the current phase when the donor is detectable. This suggests that there are very likely many Be + sdO/B binaries within the sample of Be stars from which we selected HD 15124 that have thus-far undiscovered sdO/B companions. 




We detected $N_{\rm detected}=1$ bloated stripped companions among $N_{\rm Be}=297$ Be stars. From this, we can infer $f_{\rm detectable}$, the fraction of Be stars with bloated stripped companions  detectable with our search. From Bayes' theorem, 

\begin{equation}
    \label{eq:bayes}
    p\left(f_{{\rm detectable}}|N_{{\rm detected}}\right)\propto p\left(N_{{\rm detected}}|f_{{\rm detectable}}\right)p\left(f_{{\rm detectable}}\right).
\end{equation}

Here $p\left(N_{{\rm detected}}|f_{{\rm detectable}}\right)$ is the likelihood of detecting $N_{\rm detected}$ stripped companions given $f_{\rm detectable}$. For a Poisson processes, this is given by

\begin{equation}
    \label{eq:likelihood}
    p\left(N_{{\rm detected}}|f_{{\rm detectable}}\right)=\frac{\left(f_{{\rm detectable}}N_{{\rm Be}}\right)^{N_{{\rm detected}}}}{N_{{\rm detected}}!}e^{-f_{{\rm detectable}}N_{{\rm Be}}},
\end{equation}
For the prior, $p\left(f_{{\rm detectable}}\right)$, we assume a uniform distribution between 0 and 1. The proportionality factor in Equation~\ref{eq:bayes} is then set by the normalization condition. 

To relate $f_{\rm detectable}$ to $f_{\rm stripped}$, the total fraction of Be stars that had or have stripped companions, we need an estimate of $\tau_{\rm Be}$, the lifetime of the Be phase, and $\tau_{\rm detectable}$, the lifetime of the phase during which a typical companion would have been detectable with APOGEE. For simplicity, we assume $\tau_{\rm Be} = 50\,\rm Myr$ and $\tau_{\rm detectable}= 1\,{\rm Myr}$, as suggested by our evolutionary models for HD 15124, so that 

\begin{equation}
\label{eq:lifetimes}
    \frac{f_{{\rm stripped}}}{f_{{\rm detectable}}}=\frac{\tau_{{\rm Be}}}{\tau_{{\rm detectable}}}=50.
\end{equation}

\begin{figure}
    \centering
    \includegraphics[width=\columnwidth]{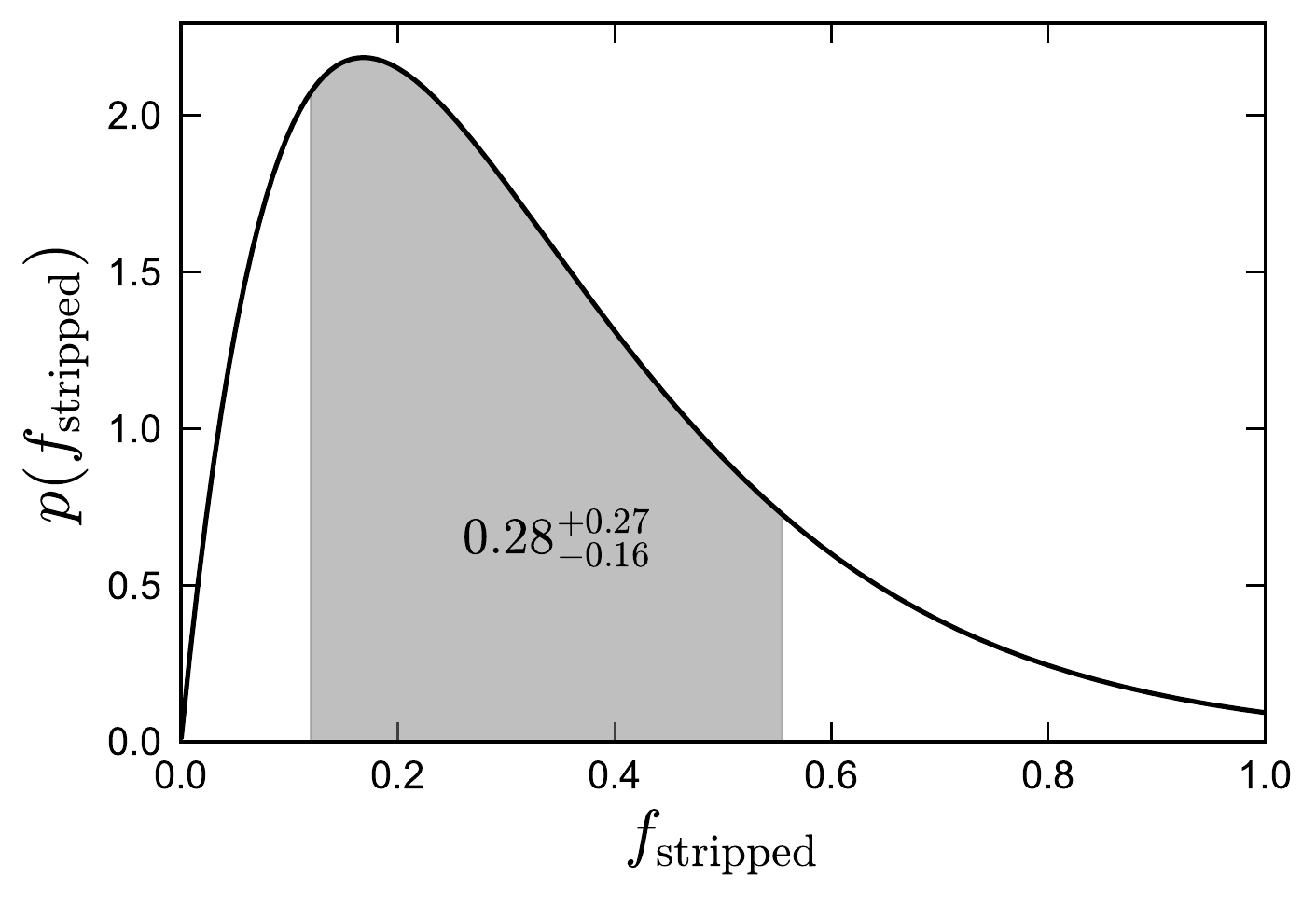}
    \caption{Constraint on the fraction of Be stars that have gone through a companion-stripping stage similar to HD 15124. This is inferred from the fact that we detected one stripped companion out of 297 Be stars and the timescale over which the companion is detectable with APOGEE spectra is $\approx 50$ times shorter than the subsequent Be star lifetime. }
    \label{fig:fstripped}
\end{figure}

Figure~\ref{fig:fstripped} shows the resulting posterior distribution of $f_{\rm stripped}$. The distribution is broad -- due entirely to Poisson uncertainties  -- but suggest that a significant fraction of Be stars have stripped companions. The median and middle 68\% of the distribution is $f_{\rm stripped} = 0.28_{-0.16}^{+0.27}$; i.e., discovery of one stripped companion implies that (10-60)\,\% of Be stars have previously had a bloated stripped companion. In most cases, these companions should survive as faint white dwarf, sdO/B, and neutron star companions.

We note that HD 15124 is in the brightest 20\% of the Be stars in initial sample (Figure~\ref{fig:cmd_selection}), raising the question whether there could be additional similar stripped companions in the sample that were missed. If we had detected a 2nd similar object, the above constraint would change to $f_{\rm stripped} = 0.45_{-0.22}^{+0.33}$. However, the stripped companion in HD 15124 is not near the limits of our search's  sensitivity (Appendix~\ref{sec:sensitivity_appendix}).   

This estimate relied on numerous approximations. Perhaps most significantly, our presumed lifetimes $\tau_{\rm Be}$ and $\tau_{\rm detectable}$ are based on models for HD 15124 and will vary somewhat with the properties of individual Be stars. More detailed calculations have been carried out \citep[e.g.][]{Pols1991, Raguzova2001, deMink2013, Schootemeijer2018, Hastings2021}. Once more bloated companions are discovered, detailed comparisons with these models will become possible. 

Our procedure for selecting Be stars and identifying candidates for having a binary companion introduces some additional systematic uncertainty. Because the parent sample within which to search for Be stars began with a CMD-based selection that only includes $\approx$ 85\% of Be stars (Section~\ref{sec:Be_search}), some Be stars observed by APOGEE never enter our sample. This will not bias our results if the stars that are excluded have similar binary properties to the rest of the sample, but it could lead to an underestimate of $f_{\rm stripped}$ if e.g. later-type Be stars or young Be stars in highly reddened regions were more likely to have binary companions. Be stars with sufficiently cool and luminous companions could also be preferentially excluded from the sample if these companions moved the unresolved source redward of our CMD search region. However, our MESA calculations predict that the stripped companions have low enough luminosity during most of the mass transfer process that this effect is minor (Figure~\ref{fig:mesa_evol}). At present, the Poisson uncertainty on $N_{\rm detected}=1$ is large enough to dominate over other sources of uncertainty. 

Our results are inconsistent at the $\approx$2 sigma level with a scenario in which {\it all} Be stars form via stable mass transfer in Algol-type systems. Not all binary evolution scenarios that could form Be stars go through an Algol-type phase. For example, some Be stars may form through binary mergers \citep[e.g.][]{deMink2013, Gies2021}. This channel could produce single Be stars. 

\section{Conclusions}
\label{sec:conclusions}
Motivated by suggestions that many Be stars are spun-up by mass transfer, we searched the spectra of Be stars from the APOGEE survey for stripped, slowly rotating companions. From a parent sample of 297 Be star candidates, we identified one promising binary, HD 15124.

The system contains a main-sequence B star with a circumstellar disk (the ``Be'' component), and a warm, bloated subgiant companion (the ``donor''). The subgiant nearly or completely fills its Roche lobe, suggesting that there is ongoing mass transfer. The Be star rotates at about 60\% of its critical rotation velocity, faster than most normal B stars, but slower than most classical Be stars. We thus propose that its circumstellar disk is an accretion disk. Ongoing mass transfer is expected to continue to spin up the star and widen the orbit, such that the accretor will become a classical Be star when mass transfer ends. HD 15124 is one of a small number of well-characterized mass transfer binaries in this state \citep[e.g.][their Table 7]{Harmanec2015}; to our knowledge, it is the first to be identified through a systematic search of B stars from a survey with a modelable selection function, enabling an estimate of occurrence rates. 

Although it satisfies the functional definition of a Be star -- it is a B star with emission lines -- we stress that the accreting star in HD 15124 is in a different evolutionary state from {\it classical} Be stars \citep[e.g.][]{Rivinius2013}, whose emission owes to episodic mass loss through a viscous decretion disk. The emission lines currently observed in HD 15124 originate in an accretion disk, which will disappear once mass transfer ends. The accreting star is, however, being spun up by mass transfer, and our modeling assumes that when mass transfer ends, it will become a classical Be star.

Modeling the system's light curves (Figures~\ref{fig:light_curves} and~\ref{fig:kelt}), high-resolution spectra (Figure~\ref{fig:apogee_cutout},~\ref{fig:pepsi},~\ref{fig:vturb}, and~\ref{fig:abundances}), radial velocities (Figure~\ref{fig:rvs}), spectral energy distribution (Figure~\ref{fig:fluxed}, and~\ref{fig:sed}), and astrometry allows us to place tight constraints on the binary's physical parameters and chemical abundances (Table~\ref{tab:system}). We interpret these using MESA binary evolution calculations (Figure~\ref{fig:mesa_evol},~\ref{fig:mesa_zoom},~\ref{fig:mesa_abun}, and~\ref{fig:mesa_profiles}), which suggest the system was born as a $\sim (4+3)\,M_{\odot}$ binary in which the subgiant was initially the more massive star. Our main results are as follows.


\begin{enumerate}
    \item {\it Observational data}: HD 15124 is a double-lined binary. The narrow-lined donor contributes $\approx 20\%$ of the optical flux, while the broad-lined Be star contributes $\approx 80\%$ (Figure~\ref{fig:fluxed}). The circumstellar disk contributes only a few percent of the optical flux, but its contributions become increasingly important at longer wavelengths (Figure~\ref{fig:sed}), and it contributes significant emission in hydrogen and helium lines (Figure~\ref{fig:pepsi}). The radial velocities of the donor are well-fit by a circular orbit with $P_{\rm orb}=5.47$\,days and projected RV semi-amplitude of $\approx 74\,\rm km\,s^{-1}$ (Figure~\ref{fig:rvs}). The light curve shows photometric variability dominated by a reflection effect and ellipsoidal variability of the donor (Figures~\ref{fig:light_curves} and~\ref{fig:kelt}). There is also variability on longer and shorter timescales, likely driven by turbulence and changes in the structure of the disk.
    \item {\it Physical parameters}: We constrain the mass and radius of the Be star to be $M_{\rm Be}=5.3\pm 0.6$ and $R_{\rm Be}=4.1\pm 0.2 R_{\odot}$, and for the donor, $M_{\rm donor}=0.92\pm 0.22$ and $R_{\rm donor}=5.82\pm 0.45 R_{\odot}$ (Table~\ref{tab:system}). The Be star is close to the main sequence (Figure~\ref{fig:be_star_teff_logg}), while the donor is clearly evolved. The inclination is $i\approx 23$ deg (close to face-on). The donor appears to be tidally synchronized, while the Be star rotates at $\approx$60\% of its critical velocity. Coupled with the fact that there is ongoing mass transfer, this suggests the emission lines originate from an accretion disk, not a decretion disk.
    \item {\it Abundances}: The surface abundances of both stars in HD 15124 are quite unusual. Helium and nitrogen are enhanced by factors of 4 and 6, while carbon and oxygen are depleted by factors of 200 and 6 (Figure~\ref{fig:abundances}). Such abundances are expected within the CNO-burning core of a few-solar mass star, where hydrogen has been converted to helium and CNO burning equilibrium is reached with most of the carbon converted to nitrogen. The fact that we observe these abundances on the stellar surface suggests that what is today the surface was once inside the convective core of the donor, and that this CNO-processed material has also been deposited on the surface of the Be star.
    
    \item  {\it Evolutionary history}: We compared the observed properties of HD 15124 to binary evolution models from the BPASS library as well as bespoke models calculated with MESA. The models that most closely match HD 15124 begin with primary and secondary masses near $4\,M_{\odot}$ and $3\,M_{\odot}$ and periods of a few days. The initially more massive star overflows its Roche lobe shortly before or after the end of its main-sequence evolution, initiating a phase of thermal-timescale mass transfer. It evolves up the giant branch as it continues to lose its envelope, passing through the currently observed phase of HD 15124 when the mass transfer rate is around $\dot{M}\sim 10^{-7}\,M_{\odot}\,\rm yr^{-1}$ (Figures~\ref{fig:mesa_evol} and~\ref{fig:mesa_zoom}). In order to match the observed short orbital period, mass transfer is required to have been fairly non-conservative, with significant loss of angular momentum after Roche lobe overflow. 
    
    Such models are predicted to terminate mass transfer within 1-2 Myr, when almost all of the envelope has been stripped and only a helium core with $\sim 0.1\,M_{\odot}$ of hydrogen envelope remains. The donor is then predicted to contract to become a hot, compact core helium burning extreme horizontal branch (or ``sdO/B'') star, where it will be a nearly invisible companion to a rapidly rotating Be star spun up by accretion. This suggests that HD 15124 is a progenitor system for the population of Be + sdO/B binaries discovered in the UV \citep[e.g.][]{Wang2021}. 
    
    \item  {\it Implications for the origin of Be stars}: Discovery of one Be + stripped star progenitor in a short-lived phase implies the existence of a larger population in the Be + sOB phase, when the stripped companion can easily escape detection. We infer that a fraction $f_{\rm stripped} = 0.28_{-0.16}^{+0.27}$ of Be stars have stripped companions (Figure~\ref{fig:fstripped}). This is broadly consistent with previous estimates from similar binaries containing slightly more evolved stripped stars \citep[e.g.][]{El-Badry2021}. The fact that HD 15124 was discovered through a systematic search of a well-defined parent sample of Be stars makes this constraint more straightforwardly interpretable than the one inferred by \citet{El-Badry2021}. The uncertainties remain large. Joint modeling of bloated and sdO/B companions to Be stars -- many more of which will be within the reach of proposed future missions \citep[e.g.][]{Jones2021, Kulkarni2021} -- will continue to solidify our understanding of the importance of mass transfer in the formation of Be stars.
 \end{enumerate}

\section*{Acknowledgements}
We thank the referee for helpful comments. We are also  grateful to David Whelan and Julia Bodensteiner for feedback on an early version of the manuscript, to Nathan Sandford and Yuan-Sen Ting for help generating model spectra, and to Robert Kurucz for making his codes publicly available. HWR acknowledges support from the GIF grant 1498.
Funding for the Sloan Digital Sky Survey IV has been provided by the Alfred P. Sloan Foundation, the U.S.  Department of Energy Office of Science, and the Participating Institutions. 
JL-B acknowledges support from FAPESP (grant 2017/23731-1).

\section*{Data Availability}
Data associated with this paper are available upon request from the corresponding author.



\bibliographystyle{mnras}



\appendix

\section{Radial velocities}
\label{sec:rv_tables}
Radial velocities from the APOGEE and NRES spectra are reported in Table~\ref{tab:rvs}.

\begin{table}
\caption{Radial velocities of the donor}
\label{tab:rvs}
\begin{tabular}{lccc}
HJD & phase & RV\,[$\rm km\,s^{-1}$] & instrument \\ 
\hline
2457681.8170 & 0.471 & $-25.58 \pm 0.96 $ & APOGEE  \\
2457734.6239 & 0.127 & $-65.01 \pm 2.14 $ & APOGEE  \\
2457762.5615 & 0.235 & $-84.82 \pm 2.16 $ & APOGEE  \\
2457765.5750 & 0.786 & $60.56 \pm 1.50 $ & APOGEE  \\
2458007.8932 & 0.091 & $-52.27 \pm 1.85 $ & APOGEE  \\
2458010.8925 & 0.640 & $45.19 \pm 1.36 $ & APOGEE  \\
2458039.8325 & 0.931 & $20.73 \pm 1.52 $ & APOGEE  \\
2458056.7598 & 0.026 & $-22.95 \pm 0.71 $ & APOGEE  \\
2458061.8121 & 0.950 & $12.03 \pm 1.40 $ & APOGEE  \\
2458064.7898 & 0.494 & $-11.22 \pm 1.21 $ & APOGEE  \\
2458084.6901 & 0.133 & $-67.23 \pm 1.62 $ & APOGEE  \\
2459247.1871 & 0.685 & $56.96 \pm 4.90 $ & NRES  \\
2459249.2283 & 0.058 & $-45.00 \pm 12.85 $ & NRES  \\
2459253.2535 & 0.794 & $60.52 \pm 5.34 $ & NRES  \\
2459256.2137 & 0.335 & $-66.62 \pm 7.73 $ & NRES  \\
2459257.1976 & 0.515 & $-1.80 \pm 6.08 $ & NRES  \\
2459259.1880 & 0.879 & $38.16 \pm 7.40 $ & NRES  \\
2459260.2069 & 0.065 & $-41.22 \pm 5.12 $ & NRES  \\
2459260.2367 & 0.071 & $-44.74 \pm 4.38 $ & NRES  \\
2459260.2710 & 0.077 & $-47.33 \pm 10.43 $ & NRES  \\
2459267.1907 & 0.342 & $-62.45 \pm 6.41 $ & NRES  \\
2459267.2539 & 0.354 & $-62.35 \pm 12.45 $ & NRES  \\
2459267.2797 & 0.359 & $-66.26 \pm 8.38 $ & NRES  \\
2459269.2240 & 0.714 & $55.56 \pm 6.80 $ & NRES  \\
2459271.2015 & 0.076 & $-46.64 \pm 2.12 $ & NRES  \\
2459271.2205 & 0.079 & $-54.62 \pm 14.74 $ & NRES  \\
2459271.2351 & 0.082 & $-45.89 \pm 3.48 $ & NRES  \\
2459272.2528 & 0.268 & $-83.40 \pm 14.82 $ & NRES  \\
2459272.2650 & 0.270 & $-78.63 \pm 3.07 $ & NRES  \\
2459274.1936 & 0.623 & $41.22 \pm 4.64 $ & NRES  \\
2459274.2129 & 0.626 & $43.83 \pm 5.07 $ & NRES  \\
2459274.2531 & 0.634 & $44.11 \pm 3.95 $ & NRES  \\
2459274.2650 & 0.636 & $45.92 \pm 1.05 $ & NRES  \\
2459275.2589 & 0.818 & $52.56 \pm 3.45 $ & NRES  \\
2459276.2598 & 0.001 & $-13.21 \pm 4.20 $ & NRES  \\
2459279.2124 & 0.540 & $9.59 \pm 2.41 $ & NRES  \\
2459279.2617 & 0.549 & $13.05 \pm 4.13 $ & NRES  \\
2459280.1959 & 0.720 & $58.42 \pm 2.89 $ & NRES  \\
2459280.2221 & 0.725 & $58.93 \pm 12.30 $ & NRES  \\
2459281.2054 & 0.905 & $28.50 \pm 5.17 $ & NRES  \\
2459282.2253 & 0.091 & $-49.22 \pm 8.87 $ & NRES  \\
\hline
\end{tabular}
\end{table}

\section{Search sensitivity}
\label{sec:sensitivity_appendix}
\begin{figure*}
    \centering
    \includegraphics[width=\textwidth]{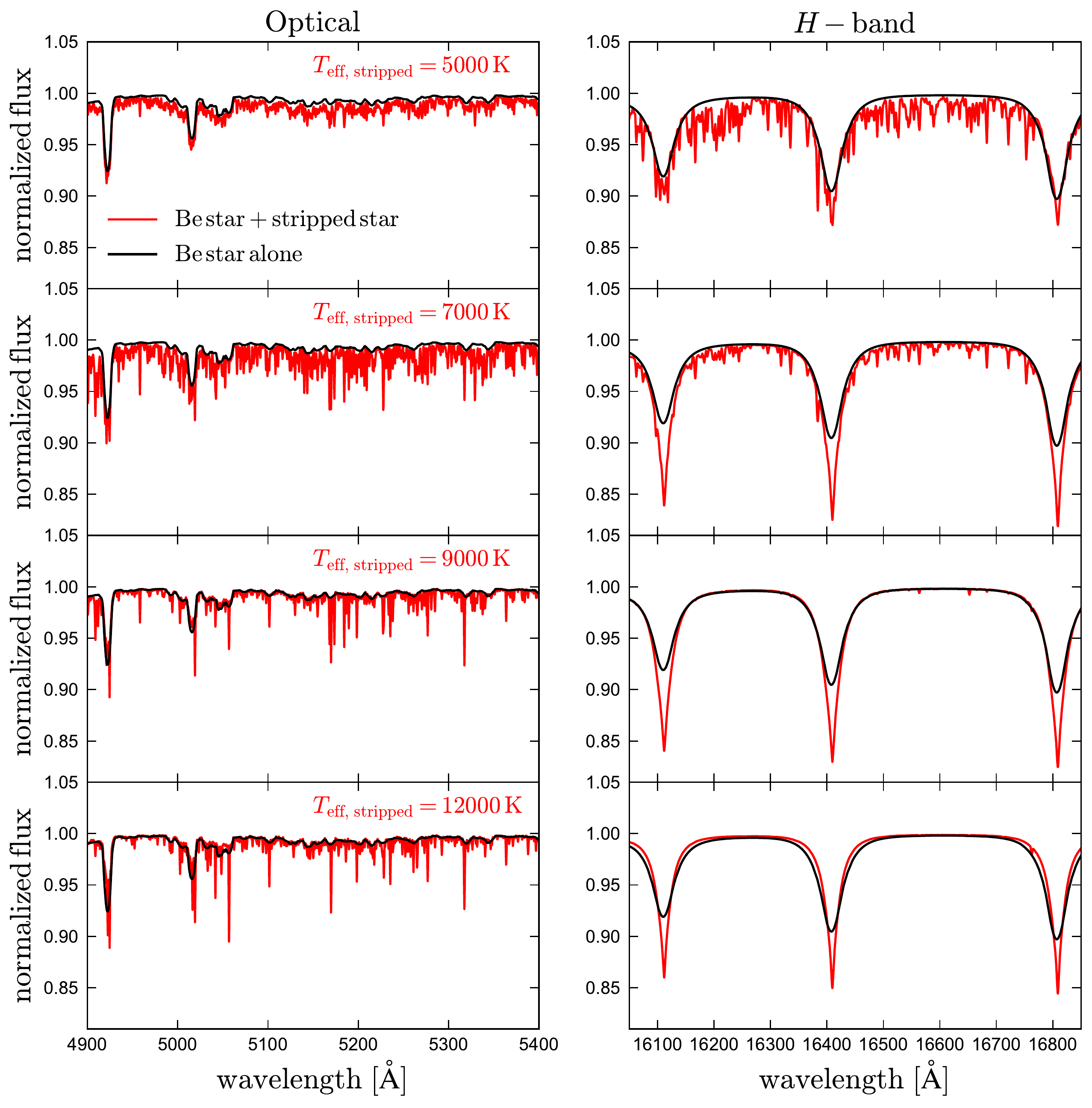}
    \caption{Simulated spectra of Be + stripped star binaries at optical (left) and APOGEE (right) wavelengths. In all panels, we assume a main-sequence Be star with $T_{\rm eff,\,Be}=18,000\,\rm K$, $R_{\rm Be}=4\,R_{\odot}$, and $v\sin i_{\rm Be}=300\,{\rm km\,s^{-1}}$. Emission lines are not included. The stripped companion has $R_{\rm stripped} = 5\,R_{\odot}$, $\log g_{\rm stripped} = 2.5$,  $v\sin i_{\rm stripped}=20\,{\rm km\,s^{-1}}$, and $T_{\rm eff,\,stripped}$ increasing from top to bottom. At APOGEE wavelengths, the narrow lines that reveal the stripped companion's presence disappear to at $T_{\rm eff,\,stripped}\gtrsim 8000\,\rm K$. In this idealized case, the Brackett lines have narrower cores when a hot stripped companion is present, but this is unlikely to be easily recognizable in practice because these will be contaminated with time-variable emission lines from the Be star. Our APOGEE search is thus only sensitive to companions with $T_{\rm eff}\lesssim 8000\,\rm K$. Hotter companions would be detectable in the optical.}
    \label{fig:sensitivity}
\end{figure*}

Figure~\ref{fig:sensitivity} shows noiseless simulated  spectra of binaries containing a rapidly rotating star (representing the Be star) and a slowly rotating companion (the stripped star) at optical and APOGEE wavelengths. Our search depends on the slowly-rotating stripped star contributing detectable narrow lines to the spectrum at APOGEE wavelengths. Such companions contribute plenty of lines when the stripped companion is relatively cool, as is the case in HD 15124. However, for $T_{\rm eff,\,stripped} \gtrsim 8,000\,\rm K$, there are very few detectable metal lines in the $H-$band. The predicted composite $H-$band spectra are still recognizably different from the predicted spectrum of a single rapidly-rotating star due to the stripped star's lower surface gravity and projected rotation velocity. However, this difference would be challenging to detect in real Be stars, whose hydrogen lines are contaminated with time-variable emission. Our search is thus sensitive primarily to companions with  $T_{\rm eff,\,stripped} \lesssim 8,000\,\rm K$; hotter companions would be more easily detectable in the optical.

%


\section{Is HD 15124 a triple?}
\label{sec:disc_triple}

HD 15124 appears in the {\it Hipparcos}-{\it Gaia} proper motion anomaly catalog \citep[][]{Kervella2019, Kervella2021}, which consists of objects whose proper motion as measured by {\it Gaia} is inconsistent with the value inferred from the difference in the {\it Hipparcos} and {\it Gaia} positions. For HD 15124, the discrepancy between the {\it Hipparcos}-{\it Gaia} eDR3 position change and the {\it Gaia} eDR3 proper motion is formally significant at the 15$\sigma$ level, with a tangential velocity anomaly of $\Delta v_{\perp }= 1.99\pm 0.13\,\rm km\,s^{-1}$. Some caution is warranted in interpreting this number, since neither the {\it Hipparcos} nor the {\it Gaia} astrometry modeled HD 15124 as an unresolved binary. However, the object's 5-day orbital period is short compared to the {\it Gaia} and {\it Hipparcos} observation windows, and the projected size of the binary's orbit in our solution is small ($\approx 0.2\,\rm mas$) compared to the parallax ($1.64 \, \rm mas$). We therefore see no strong reason to be suspicious of the reported proper motion anomaly.

Assuming the proper motion anomaly is real and due to an unresolved tertiary (following the methods described in \citet{El-Badry2021_wb}, we do not find any resolved companions in {\it Gaia} eDR3), the implied mass of the unseen companion depends on its separation. This dependence is nonlinear, because orbit-smearing effects are significant at periods shorter than the 25-year {\it Hipparcos-Gaia} epoch separation, while at periods much longer than this baseline, only a small fraction of the orbit is covered by the observations. For orbital separations of 3, 5, 10, and 30 au, \citet{Kervella2021} estimate companion masses for HD 15124 of 1.6, 0.9, 0.6, and 1.0 $M_{\odot}$. Assuming a tertiary on the main sequence, a companion with any of these masses would be much fainter than the $\sim 1000 L_{\odot}$ Be star and $\sim 100 L_{\odot}$ donor and would easily escape detection in the spectra and light curves. 

It is also worth considering again whether the Be star could be the distant tertiary to an inner binary containing the donor and an unseen companion, as is the case in $\nu$ Gem \citep{Klement2021} and NGC 2004\,\#115 \citep{El-Badry2021_ngc_2004}. The primary problem with this scenario is that it fails to explain the light curve, which shows clear modulation on the binary's 5.47 day orbital period (Figure~\ref{fig:kelt}). The shape of this modulation strongly suggests it is due to a reflection effect: the variability is too smooth to be due to eclipses, and cannot be primarily ellipsoidal variation (or its period would be half the RV period). If it is indeed a reflection effect, the companion must be hotter than the donor. Accounting for dilution from the Be star, we see no way to hide an inner companion bright enough to cause such a reflection effect. Finally, we find helium enhancement in the Be star, which is naturally explained if it has just accreted CNO-processed material from the donor, but would otherwise be unusual. 
Although the existence of a third object in HD 15124 may seem contrived, it would not be that uncommon. Indeed, the prototypical Algol-type binary -- Algol itself -- contains a low-mass tertiary, and almost 80\% of close binaries with $P_{\rm orb} < 7$ days have an outer tertiary \citep{Tokovinin2006}. About 32\% of all stars in the {\it Hipparcos} catalog have a significant proper motion anomaly.

We thus conclude that a faint tertiary is plausible, though the proper motion anomaly could also be due to systemics in the astrometry. The presence of a $0.5-2\,M_{\odot}$ companion at a few-AU separation would not significantly change any aspect of our interpretation of the inner binary.


\bsp	
\label{lastpage}
\end{document}